\documentclass[prd,twocolumn,showpacs,superscriptaddress,nofootinbib]{revtex4}
\pdfoutput=1

\usepackage{amsmath}
\usepackage{amssymb}
\usepackage{color}
\usepackage{dcolumn}
\usepackage{graphicx}
\usepackage[utf8]{inputenc}
\usepackage{latexsym}
\usepackage{times}

\newcommand{\scri}{\ensuremath{\mathcal{J}^+}}

\newcommand{\rad}{\ensuremath{\text{rad}}}
\newcommand{\tg}{\tilde\gamma}
\newcommand{\tG}{\tilde\Gamma}
\newcommand{\tA}{\tilde A}

\newcommand{\hp}{\hat{\phi}_\kappa}

\newcommand{\ie}{\textit{i.e.\ }}

\newcommand{\p}{\ensuremath{\partial}}
\DeclareMathOperator{\tr}{\ensuremath{\mathrm{tr}}}

\begin{document}

%%%%%%%%%%%%%%%%%%%%%%%%%%%%%%%%%%%%%%%%%%%%%%%%%%%%%%%%%%%%%%%%%%%%%%%%%%%%%%
%%% Title info
%%%%%%%%%%%%%%%%%%%%%%%%%%%%%%%%%%%%%%%%%%%%%%%%%%%%%%%%%%%%%%%%%%%%%%%%%%%%%%
\title{High accuracy binary black hole simulations with an extended
  wave zone}

\author{Denis Pollney}
\affiliation{
  Max-Planck-Institut f\"ur Gravitationsphysik,
  Albert-Einstein-Institut,
  Potsdam-Golm, Germany
}

\author{Christian Reisswig}
\affiliation{
  Max-Planck-Institut f\"ur Gravitationsphysik,
  Albert-Einstein-Institut,
  Potsdam-Golm, Germany
}

\author{Erik Schnetter}
\affiliation{
  Center for Computation \& Technology,
  Louisiana State University,
  Baton Rouge, LA, USA
}
\affiliation{
  Department of Physics \& Astronomy,
  Louisiana State University,
  Baton Rouge, LA, USA
}

\author{Nils Dorband}
\affiliation{
  Max-Planck-Institut f\"ur Gravitationsphysik,
  Albert-Einstein-Institut,
  Potsdam-Golm, Germany
}

\author{Peter Diener}
\affiliation{
  Department of Physics \& Astronomy,
  Louisiana State University,
  Baton Rouge, LA, USA
}
\affiliation{
  Center for Computation \& Technology,
  Louisiana State University,
  Baton Rouge, LA, USA
}

\date{2009-10-13}

\begin{abstract}
We present results from a new code for binary black hole evolutions
using the moving-puncture approach, implementing finite differences in
generalised coordinates, and allowing the spacetime to be covered with
multiple communicating non-singular coordinate patches.  Here we
consider a regular Cartesian near zone, with adapted spherical grids
covering the wave zone. The efficiencies resulting from the use of
adapted coordinates allow us to maintain sufficient grid resolution to
an artificial outer boundary location which is causally disconnected
from the measurement. For the well-studied test-case of the inspiral
of an equal-mass non-spinning binary (evolved for more than 8 orbits
before merger), we determine the phase and amplitude to numerical
accuracies better than $0.010\%$ and $0.090\%$ during inspiral, respectively,
and $0.003\%$ and $0.153\%$ during merger. The waveforms, including
the resolved higher harmonics, are convergent and can be consistently
extrapolated to $r\rightarrow\infty$ throughout the simulation,
including the merger and ringdown. Ringdown frequencies for these
modes (to $(\ell,m)=(6,6)$) match perturbative calculations to within
$0.01\%$, providing a strong confirmation that the remnant settles to
a Kerr black hole with irreducible mass $M_{\rm irr} =
0.884355\pm20\times10^{-6}$ and spin $S_f/M_f^2 = 0.686923 \pm 10\times10^{-6}$.
\end{abstract}


\maketitle

%%%%%%%%%%%%%%%%%%%%%%%%%%%%%%%%%%%%%%%%%%%%%%%%%%%%%%%%%%%%%%%%%%%%%%%%%%%%%%
%%% Text body
%%%%%%%%%%%%%%%%%%%%%%%%%%%%%%%%%%%%%%%%%%%%%%%%%%%%%%%%%%%%%%%%%%%%%%%%%%%%%%
\section{Introduction}
\label{sec:introduction}

The numerical solution of Einstein's equations has made great progress
in recent years. Orbits and mergers of binary systems of black holes
and neutron stars are now routinely published by a number of research
groups, using independent codes and
methodologies~\cite{Pretorius:2005gq, Baker:2005vv, Campanelli:2005dd,
  Scheel:2008rj}. A number of important astrophysical phenomena
associated with binary black hole mergers have been studied in some
detail. In particular, the recoil of the merger remnant has been
studied for a number of different initial data
models~\cite{Gonzalez:2006md, Gonzalez:2007hi, Campanelli:2007ew,
  Campanelli:2007cg, Herrmann:2007ac, Koppitz:2007ev, Pollney:2007ss,
  Lousto:2008dn}, and its final mass and spin has been mapped out for
fairly generic merger models involving spinning and unequal mass black
holes~\cite{Rezzolla:2007xa, Rezzolla:2007rd, Rezzolla:2007rz, Tichy:2008,
  Lousto:2007mh, Barausse:2009uz}. Since these quantities are determined by the last
few quasi-circular orbits before merger, they can be calculated to
good approximation with fairly short evolutions, and with minimal
influence of an artificial outer boundary.

Of particular topical relevance, however, 
is the construction of long
waveforms which can be used 
for gravitational-wave analysis of the binary~\cite{Reisswig:2009vc}, 
and also
to construct a family of
templates~\cite{Ajith:2007qp,Ajith:2007kx,Ajith:2007xh,Ajith:2009bn}, so to
inform and improve gravitational wave
detection algorithms. Here the requirements are particularly
challenging for numerical simulations, requiring waveforms which are
accurate in phase and amplitude over multiple cycles to allow for an
unambiguous matching to post-Newtonian waveforms at large
separation. Some recent studies have shown very promising results in
this direction for particular binary black hole
models~\cite{Baker:2006ha, Buonanno06imr, Hannam:2007ik,
  Hannam:2007wf, Damour:2007vq, Buonanno:2007pf, Damour:2008te,
  Buonanno:2009qa, Boyle:2007ft}. However, they have also highlighted
the problems associated with producing long waveforms of sufficient
accuracy.

First of all, for binaries with a larger separation,
systematic errors associated with gravitational waveform extraction at
a finite radius become more pronounced. Typically a number of
extraction radii are used, and the results extrapolated to infinite
radius (assuming such a consistent extrapolation exists given
potential issues of gauge). In order to have some confidence in the
results, the outermost ``extraction sphere'' needs to be at a large
radius, say on the order of $150-200M$ (where $M$ is the mass of the
system and sets the fiducial length scale). Even at this radius, the
amplitude of the extrapolated waveform differs significantly from the
measured waveform. Unfortunately, extracting at larger radii comes at
a computational expense. One of the standard methods in use today is
finite differencing in conjunction with ``mesh refinement'', in which
the numerical resolution is chosen based on the length scale of the
problem. A minimum number of discrete data points are required to
resolve a feature of a given size accurately, which sets a limit on
the minimum resolution which should be applied in a region. Thus,
even with mesh refinement there is a limit on the coarseness of the
grid which can be allowed in the wave-zone. For a Cartesian grid, the
number of computational points scales as $r^3$, so that requiring a
sufficient resolution to $200M$ already comes at significant expense,
and increasing this distance further becomes impractical.

An additional difficulty arises from the requirement that the outer
boundary have minimal influence on the interior evolution, since it is
(in all current binary black hole codes) an artificial boundary.  This
places an additional requirement on the size of the computational
grids, so that even outside the wave-zone region where the physics is
accurately resolved, it is conventional to place several even coarser
grids. This is done in the knowledge that the physical variables can
not be resolved in these regions, but the grids are helpful as a
numerical buffer between the outer boundary and interior
domain. Again, adding these outer zones comes at a computational
expense. The boundaries with under-resolved regions also lead to
unphysical reflections which can contaminate the solution. The problem
of increasing the grid size can be significantly reduced if, rather
than a Cartesian coordinate system, one uses a discretisation which
has a radial coordinate. Then, for a fixed angular resolution, the
number of points on the discrete grid increases simply as a linear
function of $r$, rather than the $r^3$ of the Cartesian case.  This
has two advantages. The gravitational wave-zone has spherical topology
and therefore, a numerical approximation would be most efficiently
represented by employing a spherical grid.  A further computational
motivation comes from the fact that non-synchronous mesh-refinement
(such as the Berger-Oliger algorithm) can greatly complicate the
parallelisation of an evolution scheme, and thus having many levels of
refinement clearly has an impact on the efficiency of large scale
simulations. This will become particularly relevant for the coming
generations of peta-scale machines which strongly emphasise parallel
execution (possibly over several thousand cores) over single processor
performance.

The use of spherical-polar coordinates has largely been avoided in
3-dimensional general relativity due to potential problems associated
with the coordinate singularity at the poles. Additionally, even if
regularisation were simple, the inhomogeneous areal distribution of
latitude-longitude grid points over the sphere make spherical-polar
coordinates sub-optimal.  A number of alternative coordinate systems
have been proposed and implemented for studies of black holes in
3D. The Pittsburgh null code avoids the problem of regularisation at
the poles by implementing a 2D stereographic patch
system~\cite{Bishop97b}. Cornell/Caltech have developed a multipatch
system which has been used for long binary black hole
evolutions~\cite{Scheel:2006gg, Scheel:2008rj}~\footnote{Multi-domain
  spectral methods have previously been applied to the problem of
  generating initial data for binaries in~\cite{Gourgoulhon:2000nn,
  Gourgoulhon02, Grandclement:2001ed}.}.  This code, using spectral
spatial differentiation, uses an intricate patch layout in order to
reduce the overall discretisation error. The boundary treatment
between patches is based on the transfer of characteristic variables.
A similar approach was implemented by the LSU group, for the case of
finite differences with penalty boundary
conditions~\cite{Schnetter:2006pg}, and used to successfully evolve
single perturbed black holes with a fixed
background~\cite{Dorband:2006gg} and have recently been attempted for
binary black hole systems~\cite{Pazos:2009vb}.

In this paper we describe a binary black hole evolution code based on
adapted radial coordinates in the wave zone, for generic evolution
systems. In particular, we demonstrate stable and accurate binary
black hole evolutions using BSSNOK in conjunction with this coordinate
system. The grids in the wave zone follow a prescription which was
first used by Thornburg~\cite{Thornburg:2004dv}, in which six regular
patches cover the sphere, and data at the boundaries of the patches
are filled by interpolation.  The six patch wave zone is coupled to an
interior Cartesian code, which covers the domain in which the bodies
move, and optionally allows for mesh refinement around each of the
individual bodies.  The resulting code has the advantages of making
use of established techniques for moving puncture evolutions on
Cartesian grids, while having excellent efficiency (and consequently
accuracy) in the wave zone due to the use of adapted radially-oriented
grids.

In the following sections we detail the coordinate structures which we
use. We then describe our Einstein evolution code, which is largely
based on conventional techniques common to Cartesian puncture
evolutions. Finally we perform evolutions of a binary black hole
system in order to validate the code against known results, as well as
demonstrate the ability to extract accurate waves at a large radius
with comparatively low computational cost.

%%%%%%%%%%%%%%%%%%%%%%%%%%%%%%%%%%%%%%%%%%%%%%%%%%%%%%%%%%%%%%%%%%%%%%%%%%%%%%
\section{Spacetime Discretisation}
\label{sec:multipatch}

This section describes the implementation of a generic code
infrastructure for evolving spacetimes which are covered by multiple
overlapping grid patches. A key feature of our approach is its
flexibility. It is not restricted to any particular formulation of the
Einstein equations; the mechanism for passing data between patches
(interpolation) is also formulation independent (though
characteristic~\cite{Pfeiffer:2002wt} or penalty-patch
boundaries~\cite{Carpenter1994a, Diener05b1, Pazos:2009vb} are also an
option); the size, placement and local coordinates of individual
patches are completely adaptable, provided that there is sufficient
overlap between neighbours to transfer boundary data. Further, each
patch is a locally Cartesian grid with the ability to perform
mesh-refinement to better resolve localised steep gradients, if
necessary. The particular application demonstrated in this paper is to
provide a more efficient covering of the wave-zone of an isolated
binary black hole inspiral.

At the same time, we would like to take advantage of the fact that
black hole evolutions via the ``moving puncture'' approach are well
established as a simple and effective method for stably evolving black
hole spacetimes~\cite{Baker:2005vv, Campanelli:2005dd}. By this
method, gauge conditions are applied to prevent the spacetime from
reaching the curvature singularity, so that an artificial boundary is
not required within the horizons~\cite{Hannam:2006vv}. The usual
approach is to discretise using Cartesian grids which cover the black
holes with an appropriate resolution, without special treatment or
boundary conditions for the black hole interiors, relying rather on
the causal structure of the evolution system to prevent error modes
from emerging~\cite{Brown:2009ki}.  The Cartesian grids are extended
to cover the wave zone (at reduced resolution for the sake of
efficiency), extending to a cubical grid outer boundary where an
artificial condition is applied.

A principal difficulty faced by this method is that the discretisation
is not well suited to model radial waves at large radii. In order to
resolve the wave profile, a certain minimum radial resolution is
required and must be maintained as the wave propagates to large
radii. The angular resolution, however, can remain fixed -- if a wave
is resolved at a certain angular resolution as small radii, then it is
unlikely to develop significant angular features as it propagates to
large distances from the isolated source. Cartesian grids with fixed
spacing, however, resolve spheres with an angular resolution which
scales according to $r^2$. Thus, to maintain a given required radial
resolution, the angular directions become extremely over-resolved at
large radii, and this comes at a large computational cost. Namely, for
a Cartesian grid to extend in size or increase it's resolution by a
factor $n$, the cost in memory and number of computations per timestep
increase by $n^3$, while for a radial grid with fixed angular
resolution, the increase is linear, $n$~\footnote{Note that the
  Courant limit introduces an additional factor of $n$ in each case
  due to the requirement of a reduced timestep with increasing
  resolution.}.

For the near-zone, in the neighbourhood of the orbits of the individual
bodies, the geometrical situation is not as straightforward, since
there is no clearly defined radial propagation direction between the
bodies. If the local geometry is reasonably well known (for instance,
the location of horizon surfaces), adapted coordinates can also be
considered in this regime. The technical requirements of such
coordinate systems can, however, be high. Since the bodies are moving,
the grids must move with them, or dynamical gauges chosen such that
the bodies remain in place relative to the numerical
coordinates. Potential problems arise from the coordinate singularity
if the grids are extended to $r=0$, as is the case with the standard
puncture approach. Thus, in the near-zone, Cartesian coordinates can
provide significant simplification to the overall infrastructure
requirements, while the previously mentioned drawbacks of Cartesian
coordinates are less prevalent, as it is useful to have homogeneous
resolution in each direction in situations where there is no obvious
symmetry.

The evolution code which we have constructed for the purpose of
modelling waveforms from an isolated system is based on a hybrid
approach, involving a Cartesian mesh-refined region covering the near
zone in which the bodies orbit, and a set of adapted radial grids
which efficiently cover the wave zone. The overall structure is
illustrated in Fig.~\ref{fig:7-patch-with-ghost-zones} (top), which
shows an equatorial slice of the numerical grid. Computations on each
local patch are carried out in a globally Cartesian coordinate
system. In the particular implementation considered here, the grids
overlap by some distance so that data at the boundaries between each
local coordinate patch can be communicated by interpolation from
neighbouring patches. The resulting code is both efficient, but also
simple in structure and able to take advantage of well established
methods for evolving moving puncture black holes. If suitable
interpolation methods are used, then such a system can also be used
for solutions with discontinuities and shocks as are present in
hydrodynamics.

The code has been implemented within the \texttt{Cactus}
framework~\cite{Goodale02a, cactusweb1} via extensions to the
\texttt{Carpet} driver~\cite{Schnetter-etal-03b, Schnetter06a,
  carpetweb}, which handles parallelisation via domain decomposition of
grids over processors, as well as providing the required interpolation
operators for boundary communication and analysis tools.

\subsection{Coordinate systems}
\label{sec:coords}

\begin{figure}[ht!]
  \begin{center}
    \includegraphics[width=\linewidth,trim=220 100 60 8,clip=true]
      {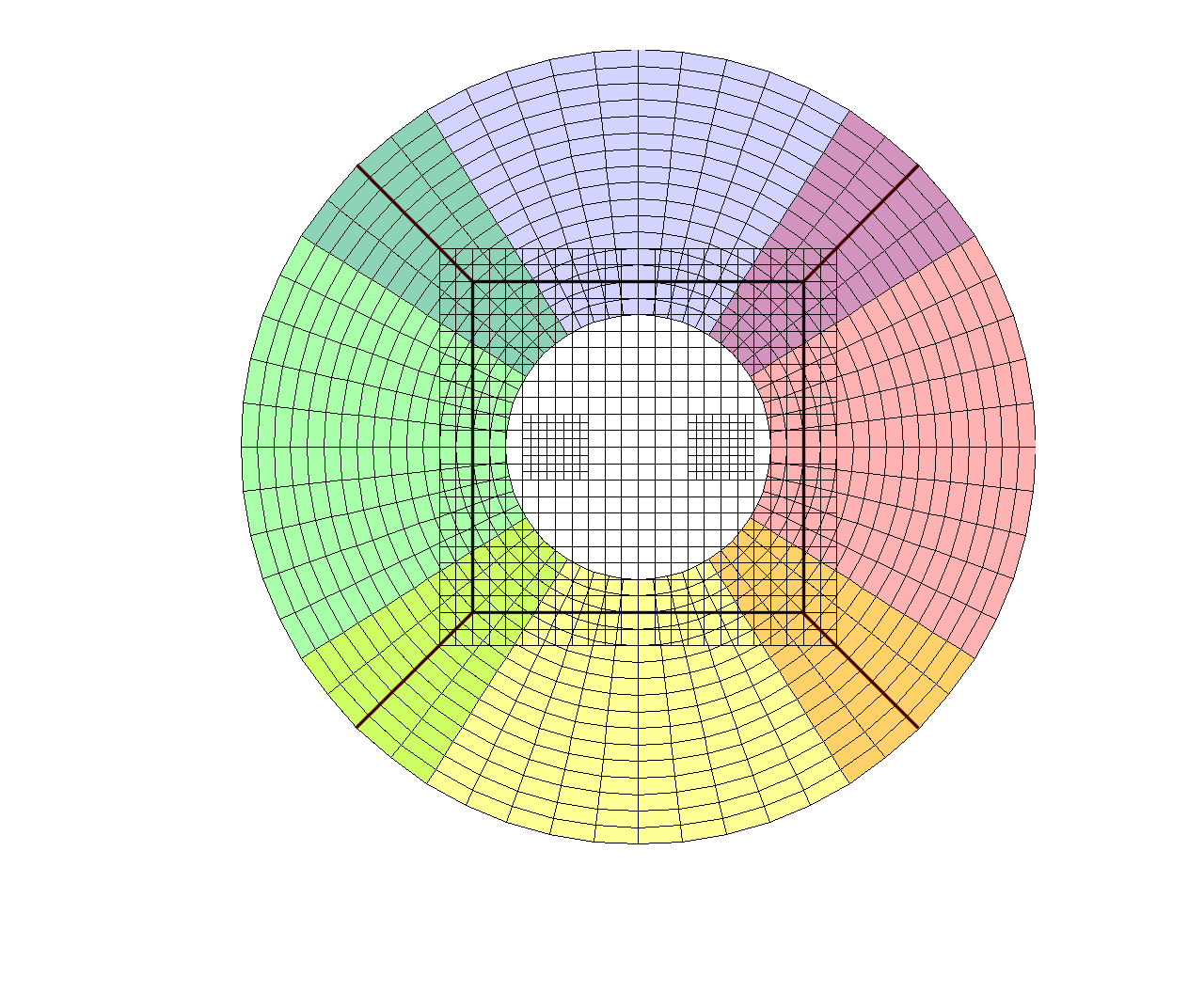}  %grids1_grey}
    \includegraphics[width=\linewidth,trim=100 0 110 0,clip=true]
      {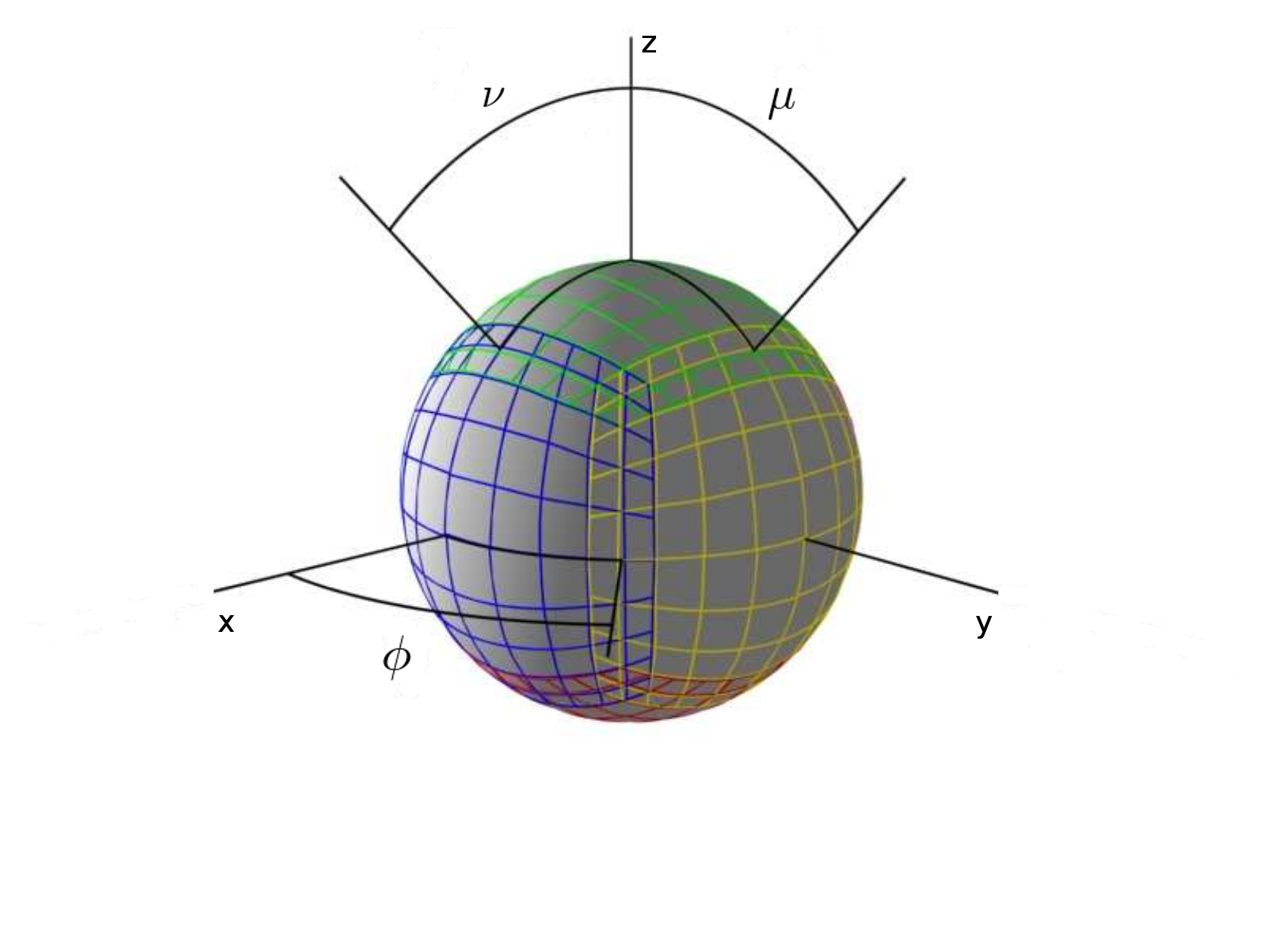}
  \end{center}
  \vspace{-2cm}
  \caption{A schematic view of the $z=0$ slice of the grid setup that
    is used. The upper plot shows the central Cartesian grid
    surrounded by six ``inflated-cube'' patches (the four equatorial
    patches are shown, shaded). The boundaries of the nominal grids owned
    by each patch are indicated by thick lines.  The lower plot
    shows an $r=\textrm{constant}$ surface of the exterior patches,
    indicating their local coordinate lines.}
  \label{fig:7-patch-with-ghost-zones}
\end{figure}

The configuration displayed in Fig.~\ref{fig:7-patch-with-ghost-zones}
consists of seven local coordinate patches: an interior Cartesian
grid, and six outer patches corresponding to the faces of the interior
cube. Each patch consists of a uniformly spaced (in local coordinates)
grid which can be refined (though in practise we will only use this
feature for the interior grid). The outer patches have a local
coordinate system $(\rho,\sigma, R)$ corresponding to the ``inflated
cube'' coordinates which have previously been used in relativity for
single black hole evolutions~\cite{Thornburg:2004dv}, and are
displayed in the lower plot of
Fig.~\ref{fig:7-patch-with-ghost-zones}.  The local angular
coordinates $(\rho,\sigma)$ range over
$(-\pi/4,+\pi/4)\times(-\pi/4,+\pi/4)$ and can be related to global
angular coordinates $(\mu,\nu,\phi)$ which are given by
\begin{subequations}
\begin{align}
  \mu  \equiv \text{rotation angle about the x-axis} &= \arctan (y/z), \\
  \nu  \equiv \text{rotation angle about the y-axis} &= \arctan (x/z), \\
  \phi \equiv \text{rotation angle about the z-axis} &= \arctan (y/x).
\end{align}
\end{subequations}
For example, on the $+z$ patch, the
mapping between the local $(\rho,\sigma,R)$ and Cartesian $(x,y,z)$
coordinates is given by:
\begin{subequations}
\begin{align}
  \rho \equiv \nu &= \arctan (x/z), \\
  \sigma \equiv \mu &= \arctan (y/z), \\
  R &= f(r),
\end{align}
\end{subequations}
with appropriate rotations for each of the other cube faces, and where
$r=\sqrt{x^2 + y^2 + z^2}$.  As emphasised by
Thornburg~\cite{Thornburg:2004dv}, in addition to
avoiding pathologies associated with the axis of standard spherical
polar coordinates, this choice of local coordinates has a number of
advantages. In particular, the angular coordinates on neighbouring
patches align so that interpolation is only 1-dimensional, in a line
parallel to the face of the patch. This potentially improves the
efficiency of the interpolation operation as well as the accuracy.
The coordinates also cover the sphere more uniformly than, say, a
stereographic 2-patch system.

The local radial coordinate, $R$, is determined as a function of the
global coordinate radius, $r$. We can use this degree of coordinate
freedom to increase the physical (global) extent of the wave-zone
grids, at the cost of some spatial resolution. In practise, we use a
function of the form
\begin{subequations}
\begin{equation}
  f(r) = A (r - r_0) + B \sqrt{1+(r-r_0)^2/\epsilon},
\end{equation}
with
\begin{equation}
  R = f(r) - f(0).
\end{equation}
\label{eq:radial_stretch}
\end{subequations}
in order to transition between two almost constant resolutions (determined
by the parameters $A$ and $B$) over a region whose width is determined
by $\epsilon$, centred at $r_0$.

\begin{figure}
  \begin{center}
    \includegraphics[width=\linewidth,trim=0 50 0 30]{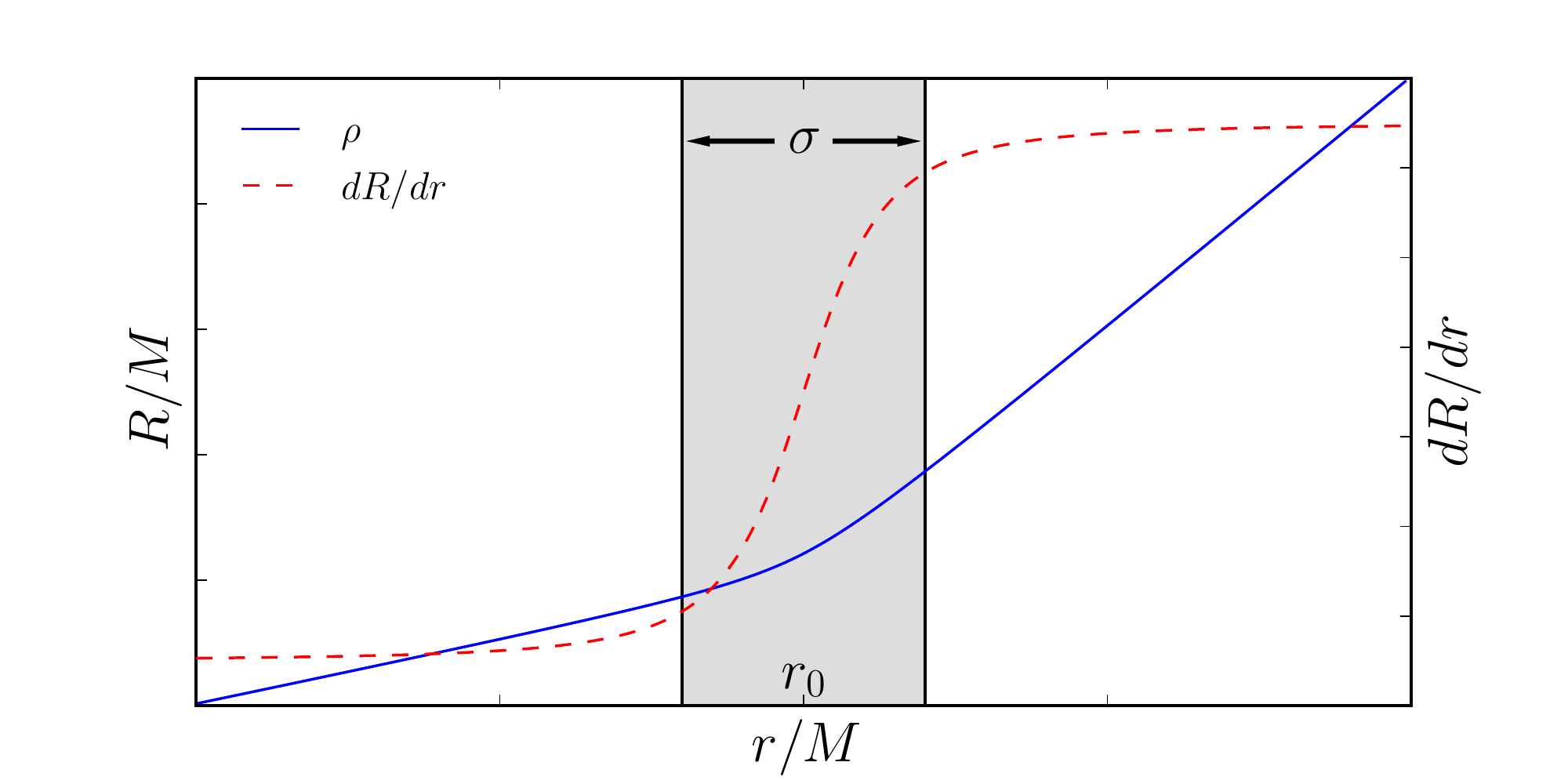}
  \end{center}
  \caption{The local radial coordinate, $R$ (solid line), can be
    stretched as a function of the global coordinate, $r$, in order to
    increase the effective size of the grid. The function specified by
    Eqs.~\eqref{eq:radial_stretch} transitions between two almost
    constant radial resolutions over a distance $\epsilon$ centred at
    $r_0$.}
  \label{fig:radial_stretch}
\end{figure}

The effect of the radial transformation is illustrated in
Fig.~\ref{fig:radial_stretch}. The coordinate $R$ is a nearly constant
rescaling of $r$ at small and large radii. The change in the scale
factor is largely confined to a transition region. Note that since we
apply the same global derivative operators (described below) to
analysis tools as are used for the the evolution, it is possible to do
analysis (e.g., measure waveforms, horizon finding) within regions
where the radial coordinate is non-uniform. The regions of
near-constant resolution are, however, useful in order to make
comparisons of measurements at different radii without the additional
complication of varying numerical error due to the underlying grid
spacing.

Data on each patch are evaluated at grid-points which are placed at
uniformly spaced points of a Cartesian grid. Thus, local derivatives
can be calculated via standard finite difference techniques. These are
then transformed to a common underlying Cartesian coordinate system by
applying an appropriate Jacobian which relates the local to global
coordinates.  That is, the global (Cartesian) coordinates, $x_i$, are
related to the local coordinates, $a_i$, by
\begin{equation}
  x_i=x_i(a_j), \qquad i,j=0,1,2.
\end{equation}
and derivatives, $\p/\p a_i$, of fields are determined using finite
differences in the regularly spaced $a_i$ coordinates, which are then
transformed using
\begin{subequations}
\begin{align}
  \frac{\p}{\p x_i} &= \left(\frac{\p a_j}{\p x_j}\right)\frac{\p}{\p a_j}, \\
  \frac{\p^2}{\p x_i \p x_j} &=
    \left(\frac{\p^2 a_k}{\p x_i \p x_j}\right) \frac{\p^2}{\p a_k^2} +
    \left(\frac{\p a_k}{\p x_i} \frac{\p a_l}{\p x_j}\right)
      \frac{\p^2}{\p a_k \p a_l},
\end{align}
\end{subequations}
in order to determine their values in the global frame.  We store and
evaluate tensor components and their evolution equations in the common
global frame, so that there is no need to apply transformations when
relating data across patch boundaries. In addition to the obvious
simplification of the inter-patch boundary treatment, this has a number
of other advantages, particularly when it comes to analysis tools
(surface finding, gravitational wave measurements, visualisation)
which may reference data on multiple patches. Since the data is
represented in the common global basis, these tools do not need to
know anything about the local grid structures or coordinates.

\subsection{Inter-patch interpolation}
\label{sec:interpolation}

\begin{figure}
  \centering
  \includegraphics[width=\linewidth]{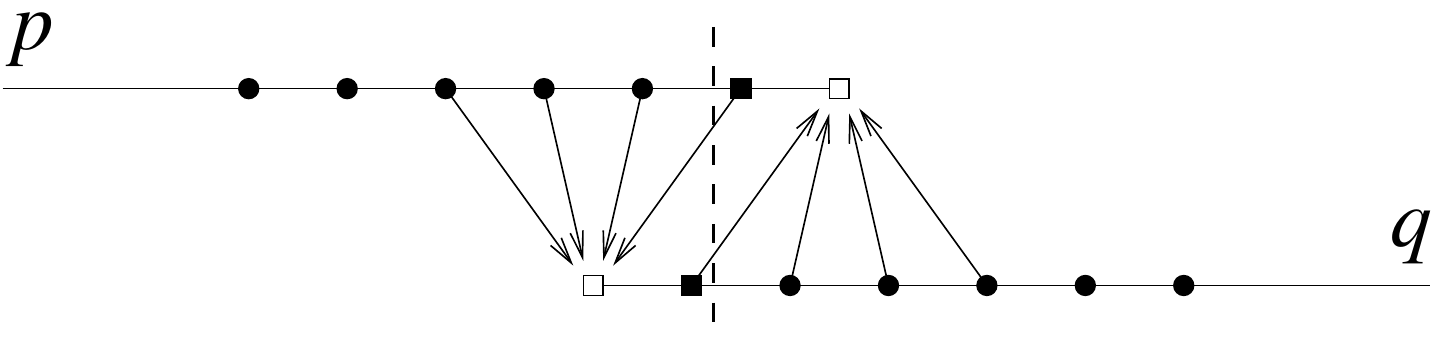}
  \caption{Schematic of interpolating patch boundaries in 1-dimension,
    assuming 4-point finite difference and interpolation stencils.
    Points in the nominal zones, $\mathcal{N}_{p,q}$, are indicated by
    filled circles, points in ghost zones, $\mathcal{G}_{p,q}$, by
    open squares, and points in overlap zones, $\mathcal{O}_{p,q}$, by
    closed squares. The vertical dotted line demarcates the boundary
    between nominal zones on each patch. Ghost points on patch $p$ are
    evaluated by centred interpolation operations from points in
    $\mathcal{S}_q$ on the overlapping patch (arrows) and \emph{vice
    versa}.}
  \label{fig:overlap}
\end{figure}

Data is communicated between patches by interpolating onto overlapping
points. Each patch, $p$, is responsible for determining the numerical
solution for a region of the spacetime. The boundaries of these
patches can overlap neighbouring patches, $q$, (and in fact must do so
for the case of the interpolating boundaries considered here),
creating regions of the spacetime which are covered by multiple
patches. We define three sets of points on a patch. The \emph{nominal}
regions, $\mathcal{N}_p$, contain the points where the numerical
solution is to be determined. The nominal regions of the patches do
not overlap, $\bigcap_p\mathcal{N}_p=\emptyset$, so that if data is
required at any point in the spacetime, it can be obtained without
ambiguity by referencing the single patch in whose nominal region it
resides. A patch, $p$, is bounded by a layer of \emph{ghost} points,
$\mathcal{G}_p$, which overlap the nominal zones of neighbouring
patches, $q$, $\mathcal{G}_p\cap\bigcup_q\mathcal{N}_q=\mathcal{G}_p$,
and are filled by interpolation. (These points are conceptually
similar to the inter-processor ghost-zones used by domain
decomposition parallelisation algorithms in order to divide grids over
processors.)  The size of these regions is determined by the width of
the finite difference stencil to be used in finite differencing the
evolution equations on the nominal grid. Finally, an additional layer
of \emph{overlap} points, $\mathcal{O}_q$, is required: i) to ensure
that the set of stencil points,
$\mathcal{S}_q\subset\mathcal{O}_q\cup\mathcal{N}_q$, used to
interpolated to the ghost zone does not itself originate from the
ghost zone of the interpolating patch,
$\mathcal{S}_q\cap\mathcal{G}_q=\emptyset$,
$\mathcal{O}_q\cap\bigcup_p\mathcal{N}_p=\mathcal{O}_q$; and ii) to
compensate for any difference in the grid spacing between the local
coordinates on the two patches. An illustration of the scheme in
1-dimension the scheme is provided in Fig.~\ref{fig:overlap}.

Note that points in
$\bigcup_q\mathcal{O}_q\subset\bigcup_p\mathcal{N}_p$ are not
interpolated, but rather are evolved in the same way as nominal grid
points within $\bigcup_p\mathcal{N}_p$. That is, in these regions
points on each grid are evolved independently, and is in principle
multi-valued. However, since the union of set of nominal points on
each patch $\bigcup_p\mathcal{N}_p$ uniquely and unambiguously covers
the entire simulation domain, \ie $\bigcap_p\mathcal{N}_p=\emptyset$,
and since the overlap regions are a subset of the nominal grid, if
data is required at a point within these overlap zones, there is
exactly one patch owing this point on its nominal grid, and it will be
returned uniquely from this patch. The differences between evolved
field values evaluated in these overlap points converge away with the
finite difference order of the evolution scheme.

The use of additional overlap points makes this inter-patch
interpolation algorithm somewhat simpler than the one implemented
by Thornburg in~\cite{Thornburg:2004dv}. That
algorithm required inter-patch boundary conditions to be applied in a
specific order to ensure that all interpolation stencils are evaluated
without using undefined grid points, and requires off-centring
interpolation stencils under certain circumstances, which is not
necessary when overlap points are added. It also relies on
the particular property of the inflated-cube coordinates which ensured
that the ghost-zones could be filled using 1-dimensional interpolation
in a direction orthogonal to the boundary. This property would be
non-trivial (and often impossible) to generalise to match arbitrary
patch boundaries, such as that between the Cartesian and radially
oriented grids of Fig.~\ref{fig:7-patch-with-ghost-zones}.

Another significant difference between Thornburg's approach and the
approach presented here is that former stores tensor components in
the patch-local frame, while we store them in the global coordinate
frame.  Evaluating components in the patch-local frame requires a
basis transformation while interpolating. This is further complicated
in the case of non-tensorial quantities (such as the $\tilde\Gamma^i$
of the BSSNOK formulation) which have quite complicated basis
transformation rules involving spatial derivatives.  Instead, we store
tensor components in the global coordinate frame, which requires no
basis transformation during inter-patch interpolations.

The number of ghost points in $\mathcal{G}_p$ can be reduced using
finite difference stencils which become lop-sided towards the
boundaries of the patch, and may provide an important optimisation
since interpolation between grids can be expensive, particularly if
processor communication is involved. However, this tends to be at the
cost of increased numerical error in the finite difference operations
towards the grid boundaries. We have generally found it preferable to
use centred stencils throughout the nominal, $\mathcal{N}_p$, and
overlap, $\mathcal{O}_p$, zones and have applied certain optimisations
to the interpolation operators as described below.  Another
optimisation can be achieved by using lower order interpolation so
that it is possible to reduce the number of overlapping points in
$\mathcal{O}_p$.

The interpolation scheme for evaluating data in ghost zones is based
on Lagrange polynomials using data from the overlapping patch. In
1-dimension, the Lagrange interpolation polynomial can be written as
\begin{subequations}
\begin{equation}
  \mathcal{L}_x[\phi](x) = \sum_{i}^{N} b_{i}(x)\, \phi_{i}\,,
\end{equation}
where the coefficients are
\begin{equation}
  b_{i}(x) =  \prod_{k\neq i}\frac{(x-x_{k})} {(x_{i}-x_{k})}\,.
  \label{eq:interp-coeff}
\end{equation}
\end{subequations}
In these expressions, $x\in\mathcal{G}_p$ is the coordinate of the interpolation point
and $\phi_i\in\mathcal{S}_q\subset\mathcal{N}_q\cup\mathcal{O}_q$ are the values at grid-points in the interpolation molecule surrounding $x$.  The number
of grid-points in the interpolation molecule, $N$, determines the
interpolation order, and interpolation of $n$-th order accuracy is
given by $N=n+1$ stencil points in the molecule.

For interpolation in $d$-dimensions, the interpolation polynomial can
be constructed as a tensor product of 1-dimensional Lagrange
interpolation polynomials along coordinate directions, $\bold{x}=(x^1,...,x^d)$:
\begin{align}
  \mathcal{L}[\phi](\bold{x}) & =
     \mathcal{L}_{x^1}[\phi](x^1)\otimes\ldots\otimes\mathcal{L}_{x^d}
       [\phi](x^d) \nonumber \\
     & = \left(\sum_{i}^{N} b_{i}(x^1)\, \phi_{i}\,,\right)
       \cdots\left(\sum_{j}^{N} c_{j}(x^d)\,\phi_{j}\right)\,.
\end{align}
Therefore, for $d$-dimensional interpolation of order $n$, one has to
determine $N^d$ neighbouring stencil points and associated
interpolation coefficients, Eq.~\eqref{eq:interp-coeff}, \emph{for each}
point in the ghost-zone of a given patch.  Most generally, full
3-dimensional interpolation is required, though in particular cases
coordinates between two patches can be constructed such that they
align locally so that only 1-dimensional interpolation is
needed. This is, for instance, the case for the overlap region
between the inflated-cube spherical patches used here (see
Fig.~\ref{fig:7-patch-with-ghost-zones}). We have
optimised the current code to automatically take advantage of this.

In order to interpolate to a point for which the coordinates $a^p_i$
given in the basis of patch $p$ are given, we need to know the patch
owning the nominal region containing this point.  For this we first
convert $a^p_i$ to the global coordinate basis $x_i$, then determine
which patch $q$ owns the corresponding nominal region $\mathcal{N}_q$, and then
convert $x_i$ to the local coordinate bases this patch $a^q_i$.  By
construction, patch $q$ has sufficient overlap points to evaluate the
interpolation stencil there:
\begin{subequations}
\begin{align}
  x_i   & := \textrm{local-to-global}_p (a^p_i)\,, \\
  q     & := \textrm{owning-patch} (x^i)\,, \\
  a^q_i & := \textrm{global-to-local}_q (x^i) \,.
\end{align}
\end{subequations}
The three operations ``$\textrm{local-to-global}$'',
``$\textrm{owning-patch}$'', and ``$\textrm{global-to-local}$'' depend
on the patch system and their local coordinate systems.

We can then find the points of patch $q$ that are closest to the
interpolation point $a^q_i$ in the local coordinates this patch.  In
order to find these points, we exploit the uniformity of the grid in
local coordinates.  The grid indices of the stencil points in a given
direction are determined via
\begin{equation}
  i \in\left\{ \text{floor}(j+k)\; \middle| \;
    j=\frac{x-x_0}{\Delta x},\;  k=-\frac{n}{2},\cdots,\frac{n}{2} \right\}\,,
\end{equation}
where $x_0$ is the origin of the local grid, $n$ is the interpolation
order, and ``floor'' denotes rounding downwards to the nearest
integer.

There remains to be determined the refinement level which contains the
region surrounding the interpolation point, as well as the processor
that owns this part of the grid.  For this purpose, an efficient
tree-search algorithm has been implemented.  In this algorithm, the
individual patches and refinement levels are defined as
``super-regions'', i.e., bounding boxes that delineate the global grid
extent before processor decomposition.  Each of these super-regions is
recursively split into smaller regions.  The leaves of the resulting
tree structure represent the individual local components of the
processor decomposition.  Locating a grid point in this tree structure
requires $O(\log n)$ operations on $n$ processors, whereas a linear
search (that would be necessary without a tree structure) would
require $O(n)$ operations.

While the corresponding tree structure is generic, the actual algorithm
used in \texttt{Carpet} splits the domain into a $kd$ tree of depth
$d$ in $d=3$ dimensions.  That is, the domain is first split
into $k$ sub-domains in the $x$ direction, each of these sub-domains
is then independently split into several in the $y$ direction, and
each of these is then split in the $z$ direction.  This leads to
cuboid sub-domains for each processor, where the sub-domains do not
overlap, and where each sub-domain can have a different shape.
\texttt{Carpet} balances the load so that each processor receives
approximately the same number of grid points, while keeping the
sub-domains' shapes as close to a cube as possible.

Our implementation pre-calculates and stores most of the above
information when
the grid structure is set up, saving a significant amount of time
during interpolation.  In particular, the following are stored:
\begin{itemize}
\item For each ghost-point, the source patch (where the interpolation
  is performed), and the local coordinates on this patch;
\item For each ghost-point, the interpolation stencil coefficients
  (\ref{eq:interp-coeff});
\item For each processor, the communication schedule specifying which
  interpolation points need to be sent to what other processor.
\end{itemize}
When the grid structure changes, for example, when a mesh-refinement
grid is moved or resized, these quantities have to be recalculated.

Altogether, the inter-patch interpolation therefore consists of
applying processor-local interpolation stencils, sending the results
to other processors, receiving results from other processors, and
storing these results in the local ghost-points.  These are all
operations requiring no look-up in complex data structures, and which
consequently execute very efficiently on modern hardware.

\subsection{Finite differencing}

Spatial derivatives are computed using standard finite difference
stencils, which have currently been implemented up to
8th-order~\cite{Diener05b1}. The stencils are centred, except for the
terms corresponding to an advection by the shift vector, of the form
$\beta^i\partial_i u$ (see Sec.~\ref{sec:evolution}, below). These
derivatives are calculated using an ``upwind'' stencil which is
shifted by one point in the direction of the shift, and of the same
order. We find that these upwind stencils provide a significant
increase in the numerical accuracy of the puncture motion at a given
resolution (see Appendix~\ref{sec:upwind}).  The particular stencils
which we use can be generated via a recursion algorithm, as outlined
in~\cite{Fornberg-1988}.

On each patch we allow the local grids to be refined in order to
increase the accuracy of computations in localised regions.  For the
application of the evolution of an isolated binary considered here, we
only refine the central Cartesian grid in the neighbourhood the
bodies. This is done using standard $2:1$ Berger-Oliger mesh
refinement techniques via the \texttt{Carpet}
infrastructure~\cite{Schnetter-etal-03b, Schnetter06a, carpetweb}. The
time step for the outer patches is taken to be the same as the coarse
grid step of the interior patch, so that no time-interpolation is
required at inter-patch boundaries.

Time integration is carried out using standard method-of-lines
integrators. We find that for the time resolution we are using, a
4th-order Runge-Kutta (RK4) method provides a good compromise between
sufficient accuracy and a low memory footprint. We set the time
resolution of the outer grids according to the timestep of the
coarsest Cartesian grid, which is limited by the Courant condition at
the specified spatial resolution. By placing the Cartesian-spherical
boundary at a small radius (and thus extending to finer Cartesian
grids) we attain a high time resolution in the wave zone, potentially
important for determining higher harmonics.

\subsection{Surface integration and harmonic decomposition}

A number of quantities of physical interest are measured by projecting
them onto closed surfaces surrounding the source. In particular,
gravitational wave measurements rely on computations on constant
coordinate spheres $S^2$, parameterized by local spherical-polar
coordinates $(\theta, \phi)$ with constant coordinate radius $r$.  In
principle, it would be possible to construct coordinates on these
2-dimensional spheres which correspond to the underlying grid
coordinates of the evolution, for instance as portrayed in the lower
figure of Fig.~\ref{fig:7-patch-with-ghost-zones}. In this case, data
can be mapped directly onto the spheres. More generally, however,
interpolation can be used to obtain data at points on the measurement
spheres, according to the procedure outlined in
Sec.~\ref{sec:interpolation}, above.

For the purpose of analysis, it is often convenient to decompose the
data on $S^2$ in terms of (spin-weighted) spherical harmonic modes,
\begin{equation}
  A_{\ell m}=\int d\Omega \sqrt{-g} A(\Omega) {}_s\bar{Y}_{\ell m}(\Omega)\,,
  \label{eq:harmonic}
\end{equation}
where $g$ is the determinant of the surface metric and $\Omega$
angular coordinates.  In standard spherical-polar coordinates
$(\theta, \phi)$,
\begin{equation}
  \sqrt{-g} = \sin^2\theta\,.
\end{equation}
The integral, Eq.~\eqref{eq:harmonic}, is solved numerically as follows.
In the spherical polar case, we can take advantage of an highly
accurate Gauss quadrature scheme which is exact for polynomials of
order up to $2N-1$, where $N$ is the number of gridpoints.  More
specifically, we use Gauss-Chebyshev quadrature.  The scheme can be
written out as
\begin{equation}
  \int d\Omega f(\Omega) = \sum_i^{N_\theta}\sum_j^{N_\phi} f_{ij} w_{j}
  + \mathcal{O}(N_\theta)\,,
  \label{eq:poly-int}
\end{equation}
where $N_\theta$ and $N_\phi$ are the number of angular gridpoints and
$w_j$ are weight functions~\cite{Driscoll94, Bateman55},
\begin{eqnarray}
  w_{j} &=& \frac{2\pi}{N_\phi} \frac{1}{N_\theta\sqrt{2\pi}} \sum_{l=0}^{N_\theta/2-1}\frac{1}{2l+1}\sin\left([2l+1]\frac{\pi j}{N}\right)\,, \nonumber \\
      & & \qquad j=0,...,N_\theta-1\,. 
\end{eqnarray}
In our implementation, the weight functions are pre-calculated for
fast surface integration.

%%%%%%%%%%%%%%%%%%%%%%%%%%%%%%%%%%%%%%%%%%%%%%%%%%%%%%%%%%%%%%%%%%%%%%%%%%%%%%
\section{Evolution system}
\label{sec:evolution}

We evolve the spacetime using a variant of the ``BSSNOK'' evolution
system~\cite{Nakamura87, Shibata95, Baumgarte99, Alcubierre99d} and a
specific set of gauges \cite{Alcubierre02a, vanMeter:2006vi}, which
have been shown to be effective at treating the coordinate
singularities of Bowen-York initial data.

The 4-geometry of a spacelike slice $\Sigma$ (with timelike normal,
$n^\alpha$) is determined by its intrinsic 3-metric, $\gamma_{ab}$ and
extrinsic curvature, $K_{ab}$, as well as a scalar lapse function,
$\alpha$, and shift vector, $\beta^a$ which determine the coordinate
propagation.  The standard BSSNOK system defines modified variables by
performing a conformal transformation on the 3-metric,
\begin{equation}
  \label{eq:def_g}
  \phi := \frac{1}{12}\ln \det \gamma_{ab}, \qquad
  \tg_{ab} := e^{-4\phi} \gamma_{ab},
\end{equation}
subject to the constraint
\begin{equation}
  \label{eq:phi_constraint}
  \det\tg_{ab} = 1,
\end{equation}
and by removing the trace of $K_{ab}$,
\begin{align}
  \label{eq:def_K}
  K & := \tr K_{ij} = g^{ij} K_{ij}, \\
  \tA_{ij} & := e^{-4\phi} (K_{ij} - \frac{1}{3}\gamma_{ij} K),
\end{align}
with the constraint
\begin{equation}
  \label{eq:A_constraint}
  \tA := \tg^{ij}\tA_{ij} = 0.
\end{equation}
Additionally, three new variables are introduced, defined in terms
of the Christoffel symbols of $\tilde{\gamma}_{ab}$ by
\begin{equation}
  \label{eq:def_Gamma}
  \tG^a := \tg^{ij}\tG^a_{ij}.
\end{equation}
In principle the $\tG^a$ can be determined from the $\tg_{ab}$, on a
slice however their introduction is key to establishing a strongly
hyperbolic (and thus stable) evolution system. In practise, only the
constraint Eq.~(\ref{eq:A_constraint}) is enforced during
evolution~\cite{Alcubierre99e}, while Eq.~(\ref{eq:phi_constraint}) and
Eq.~(\ref{eq:def_Gamma}) are simply monitored as indicators of numerical
error. Thus, the traditional BSSNOK prescription evolves the variables
\begin{equation}
  \phi,\quad \tg_{ab},\quad K,\quad \tA_{ab},\quad \tG^a,
\end{equation}
according to evolution equations which have been written down a
number of times (see~\cite{Baumgarte:2002jm, Alcubierre:2008}
reviews).

In the context of puncture evolutions, it has been noted that
alternative scalings of the conformal factor may exhibit better
numerical behaviour in the neighbourhood of the puncture as compared
with $\phi$. In particular, a variable of the form
\begin{equation}
  \hp := (\det\gamma_{ab})^{-1/\kappa},
\end{equation}
has been suggested~\cite{Campanelli:2005dd,
  Marronetti:2007wz}. In~\cite{Campanelli:2005dd}, it is noted that
certain singular terms in the evolution equations for Bowen-York
initial data can be corrected using
$\chi := \hat{\phi}_3$. Alternatively,~\cite{Marronetti:2007wz}
notes that $W := \hat{\phi}_6$ has the additional benefit of
ensuring $\gamma$ remains positive, a property which needs to be
explicitly enforced with $\chi$.

In terms of $\hp$, the BSSNOK evolution equations become:
\begin{subequations}
\begin{align}
  \p_t \hp      = & \frac{2}{\kappa} \hp \alpha K 
                  + \beta^i\p_i\hp 
                  - \frac{2}{\kappa} \hp\p_i\beta^i,
                  \label{eq:evo_phi} \\
  \p_t \tg_{ab} = & -2 \alpha \tA_{ab} 
                  + \beta^i\p_i\tg_{ab} 
                  + 2 \tg_{i(a}\p_{b)}\beta^i
                  \label{eq:evo_tg} \\
                  & - \frac{2}{3}\tg_{ab} \p_i\beta^i,
		  \nonumber \\
  \p_t K        = & -D_iD^i \alpha + \alpha (A_{ij}A^{ij} 
                  + \frac{1}{3} K^2) 
                  + \beta^i\p_iK,
                  \label{eq:evo_K} \\
  \p_t \tA_{ab} = & (\hp)^{\kappa/3} (-D_aD_b\alpha 
                  + \alpha R_{ab})^\text{TF}
                  + \beta^i\p_i\tA_{ab}
                  \label{eq:evo_tA} \\
                  & + 2\tA_{i(a}\p_{b)}\beta^i
                  - \frac{2}{3}A_{ab}\p_i\beta^i,
		  \nonumber \\
  \p_t \tG^a    = & \tg^{ij}\p_i\beta_j\beta^a 
                  + \frac{1}{3} \tg^{ai}\p_i\p_j\beta^j
                  - \tG^i\p_i\beta^a
                  \label{eq:evo_tG} \\
                  & + \frac{2}{3}\tG^a \p_i\beta^i
                  - 2\tA^{ai} \p_i\alpha
                  \nonumber \\
                  & + 2\alpha (\tG^a_{ij} \tA^{ij} 
                               - \frac{\kappa}{2} \tA^{ai}\frac{\p_i\hp}{\hp}
                               - \frac{2}{3} \tg^{ai}\p_iK),  \nonumber
\end{align}
\label{eq:bssn}
\end{subequations}
where $D_a$ is the covariant derivative determined by $\tg_{ab}$, and
``TF'' indicates that the trace-free part of the bracketed term is
used.

We have implemented the traditional $\phi$ form of the BSSNOK
evolution equations, as well as the $\chi$ and $W$ variants implicit
in the evolution system, Eqs.~\eqref{eq:bssn}.  All three evolution
systems produce stable evolutions of binary black holes, though the
$\chi$ variant requires some special treatment if, due to numerical
error particularly in the neighbourhood of the punctures,
$\hat{\phi}_3\le 0$~\cite{Brugmann:2008zz}. Generally we find that the
advection of the puncture (and thus the phase accuracy of the
simulation) exhibits lower numerical error when using the $\chi$ and $W$
variants (see Appendix~\ref{sec:conformal_factor}). Convergence
properties of physical variables (such as measured gravitational
waves, or constraints measured outside of the horizons), however, are
not affected by the choice of conformal variable.

The Einstein equations are completed by a set of four constraints
which must be satisfied on each spacelike slice:
\begin{subequations}
\begin{align}
  \label{eq:einstein_ham_constraint}
  \mathcal{H} &\equiv R^{(3)} + K^2 - K_{ij} K^{ij} = 0, \\
  \label{eq:einstein_mom_constraints}
  \mathcal{M}^a &\equiv D_i(K^{ai} - \gamma^{ai}K) = 0.
\end{align}
\end{subequations}
Again, we do not actively enforce these equations, but rather
monitor their magnitude in order to determine whether our
numerical solution is deviating from a solution to the Einstein
equations.

The gauge quantities, $\alpha$ and $\beta^a$, are evolved using the
prescriptions that have been commonly applied to BSSNOK black hole,
and particularly puncture, evolutions. For the lapse, we evolve
according to the ``$1+\log$'' condition~\cite{Bona95b},
\begin{equation}
  \partial_t \alpha - \beta^i\partial_i\alpha 
    = -2 \alpha K,
  \label{eq:one_plus_log}
\end{equation}
while the shift is evolved using the hyperbolic ``$\tG$-driver''
equation~\cite{Alcubierre02a},
\begin{subequations}
\begin{align}
  \partial_t \beta^a - \beta^i \partial_i  \beta^a & 
    = \frac{3}{4} \alpha B^a\,, \\
  \partial_t B^a - \beta^j \partial_j B^i &
    = \partial_t \tilde\Gamma^a - \beta^i \partial_i \tilde\Gamma^a
    - \eta B^a\,,
\end{align}
\end{subequations}
where $\eta$ is a parameter which acts as a (mass dependent) damping
coefficient, and is typically set to values on the order of unity for
the simulations carried out here. The advective terms in these
equations were not present in the original definitions
of~\cite{Alcubierre02a}, where co-moving coordinates were used, but
have been added following the experience of more recent studies using
moving punctures~\cite{Baker:2005vv, vanMeter:2006vi}.

\subsection{Wave extraction}

The Newman-Penrose formalism \cite{Newman62a} provides a convenient
representation for a number of radiation related quantities as
spin-weighted scalars. In particular, the curvature component
\begin{equation}
  \psi_4 \equiv -C_{\alpha\beta\gamma\delta}
    n^\alpha \bar{m}^\beta n^\gamma \bar{m}^\delta,
  \label{eq:psi4def}
\end{equation}
is defined as a particular component of the Weyl curvature,
$C_{\alpha\beta\gamma\delta}$, projected onto a given null frame,
$\{\boldsymbol{l}, \boldsymbol{n}, \boldsymbol{m},
\bar{\boldsymbol{m}}\}$. 

The identification of the Weyl scalar $\psi_4$ with the
gravitational radiation content of the spacetime is a result of
the peeling theorem \cite{Sachs61, Newman62a, Penrose:1963}, which states that in an appropriate frame
and for sufficiently smooth and asymptotically flat initial data near spatial infinity,
the $\psi_4$ component of the curvature has the slowest fall-off
with radius, $\mathcal{O}(1/r)$.

The most straight-forward way of evaluating $\psi_4$ in numerical
(Cauchy) simulations is to define an orthonormal basis in the three
space $(\hat{\boldsymbol{r}}, \hat{\boldsymbol{\theta}},
\hat{\boldsymbol{\phi}})$, centered on the Cartesian grid center and
oriented with poles along $\hat{\boldsymbol{z}}$. The normal to the
slice defines a time-like vector $\hat{\boldsymbol{t}}$, from which we
construct the null frame
\begin{equation}
\label{numframe}
   \boldsymbol{l} = \frac{1}{\sqrt{2}}(\hat{\boldsymbol{t}} - \hat{\boldsymbol{r}}),\quad
   \boldsymbol{n} = \frac{1}{\sqrt{2}}(\hat{\boldsymbol{t}} + \hat{\boldsymbol{r}}),\quad
   \boldsymbol{m} = \frac{1}{\sqrt{2}}(\hat{\boldsymbol{\theta}} - 
     {\mathrm i}\hat{\boldsymbol{\phi}}) \ .
\end{equation}
Note that in order to make the vectors $\{\boldsymbol{l},
\boldsymbol{n}, \boldsymbol{m}, \bar{\boldsymbol{m}}\}$ null,
$(\hat{\boldsymbol{r}}, \hat{\boldsymbol{\theta}},
\hat{\boldsymbol{\phi}})$ have to be orthonormal relative to the
spacetime metric. In practice, we fix $\hat{\boldsymbol{r}}$ and then
apply a Gram-Schmidt orthonormalization procedure to determine
$\hat{\boldsymbol{\theta}}$ and
$\hat{\boldsymbol{\phi}})$~\footnote{Alternative frame constructions have
  also been used, such as orthonormalising relative to one of the
  angular basis vectors~\cite{Baker:2001sf}, or omitting the
  orthonormalisation altogether~\cite{Scheel:2008rj}. We have
  generally found these modifications do not lead to significantly
  different measurements}. 
It is then possible to calculate $\psi_4$ via a reformulation of
(\ref{eq:psi4def}) in terms of the geometrical variables on the
slice~\cite{Shinkai94} via the electric and magnetic parts of the
Weyl tensor,
\begin{equation}
  \psi_4 = C_{ij} \bar{m}^i \bar{m}^j\,,  \label{eq:psi4_adm}
\end{equation}
where
\begin{equation}
  C_{ij} \equiv E_{ij}-iB_{ij} = R_{ij} - K K_{ij} + K_i{}^k K_{kj} 
    - {\rm i}\epsilon_i{}^{kl} \nabla_l K_{jk}\,. \label{eq:cij}
\end{equation}
The remaining Weyl scalars can be similarly obtained and read
\begin{subequations}
\begin{eqnarray}
\psi_3 &=& \frac{1}{\sqrt{2}}C_{ij}\bar{m}^i e_r^j\,, \\
\psi_2 &=& \frac{1}{2}C_{ij}e_r^i e_r^j\,, \\
\psi_1 &=& -\frac{1}{\sqrt{2}}C_{ij}m^i e_r^j\,,  \\
\psi_0 &=& C_{ij}m^i m^j\,, \label{eq:psi0_adm}
\end{eqnarray}
\end{subequations}
where $(e_r^j)\equiv \hat{\boldsymbol{r}}$ is the unit radial vector.

In relating $\psi_4$ to the gravitational radiation, one is limited by
the fact that the measurements have been taken at a finite radius from
the source. Local coordinate and frame effects can complicate the
interpretation of $\psi_4$. These problems can largely be alleviated
by taking measurements at several radii and performing polynomial
extrapolations to $r\rightarrow\infty$. Procedures for doing so have
been studied in~\cite{Boyle:2009vi,
  Pollney:2009MP-unpublished}. In~\cite{Pollney:2009MP-unpublished} we
have shown that if a sufficiently large outermost extrapolation radius
is used, the variation in this procedure is of the order $\Delta
A=0.03\%$ and $\Delta\phi = 0.003\,\rad$ in amplitude and phase
respectively, and is consistent throught the evolution, including
inspiral, merger and ringdown. The extrapolation error is larger than
the numerical error determined in Sec.~\ref{sec:accuracy}, below, even
if it is performed using data at $r=1000M$ distant from the source,
highlighting the need for measurements at large radii. For the
``extrapolated'' data plotted in this paper, we have performed
polynomial extrapolations as detailed
in~\cite{Pollney:2009MP-unpublished}, using the six outermost
measurements at $r=\{280M,300M,400M,500M,600M,1000M\}$.

In a companion paper~\cite{Reisswig:2009us}, we use the same dataset
to calculate $\psi_4$ directly at $\scri$ using characteristic
extraction~\cite{Bishop98b, Babiuc:2009}. Here the traditional
approach (which is gauge dependent and has a finite-radius cut-off
error) presented here is replaced by a characteristic formulation of
the Einstein equations in order to determine the fields out to future
null infinity. In this paper, we restrict ourselves to a discussion of
the numerical error inherent in the evolution procedure via the
multi-patch code, and will report in more detail on systematic
measurement errors due to finite radius effects and the characteristic
extraction procedure elsewhere~\cite{Reisswig:2009us,
  Reisswig:2009-cce-long}.

%%%%%%%%%%%%%%%%%%%%%%%%%%%%%%%%%%%%%%%%%%%%%%%%%%%%%%%%%%%%%%%%%%%%%%%%%%%%%%

\section{Code verification}
\label{sec:convergence}

\subsection{Initial data}
\label{sec:initial_data}

To demonstrate the efficacy of the infrastructure described in the
previous sections, we have carried out an evolution of the by now
well-studied case of the late-inspiral and merger of a pair of
non-spinning equal-mass black holes (see, for example,
\cite{Hannam:2009hh} for an extensive discussion of numerical results
involving this model). The particular numerical evolution which we have
carried out starts from an initial separation $d/M=11.0$ and goes
through approximately 8 orbits (a physical time of around $1360M$),
merger and ring-down. The masses of the punctures are set to
$m=0.4872$ and are initial placed on the $x$-axis with momenta
$p=(\pm0.0903, \mp 0.000728, 0)$, giving the initial slice an ADM mass
$M_{\rm ADM}=0.99051968 \pm 2\times 10^{-8}$. These initial data
parameters were determined using a post-Newtonian evolution from large
initial separation, following the procedure outlined
in~\cite{Husa:2007rh}, with the conservative part of the Hamiltonian
accurate to 3PN, and radiation-reaction to 3.5PN, and determines
orbits with a measured eccentricity of $e=0.004\pm0.0005$.

\subsection{Grid setup}

The binary black hole evolution was carried out on a 7-patch grid
structure, as described in Sec.~\ref{sec:multipatch}, incorporating a
Cartesian mesh-refined region which covers the near-zone, and six
radially oriented patches covering the wave-zone.

The inner boundary of the radial grids was placed at $r_t=35.2M$
relative to the centre of the Cartesian grid. As a general rule, this
boundary should be made as small as possible to improve efficiency in
terms of memory usage.  However other factors may make it preferable
to move it further out. In particular, since we do not perform time
interpolation at grid boundaries, the time step $dt$ of the coarsest
Cartesian grid determines the timestep of the radial grids, and thus
the wave zone.  Updates of the radial grids tend to be expensive, as
they are large, so that if $dt$ is too small, computation time may be
spent over-resolving (in time) the wave zone. Particularly if the
principle interest is in the lower order wave modes, it may be optimal
to add an additional Cartesian mesh-refinement grid with a coarser
time-step, and thus move $r_t$ outwards.

\begin{figure}
  \begin{center}
    \includegraphics[width=\linewidth,clip,trim=0 0 0 0]{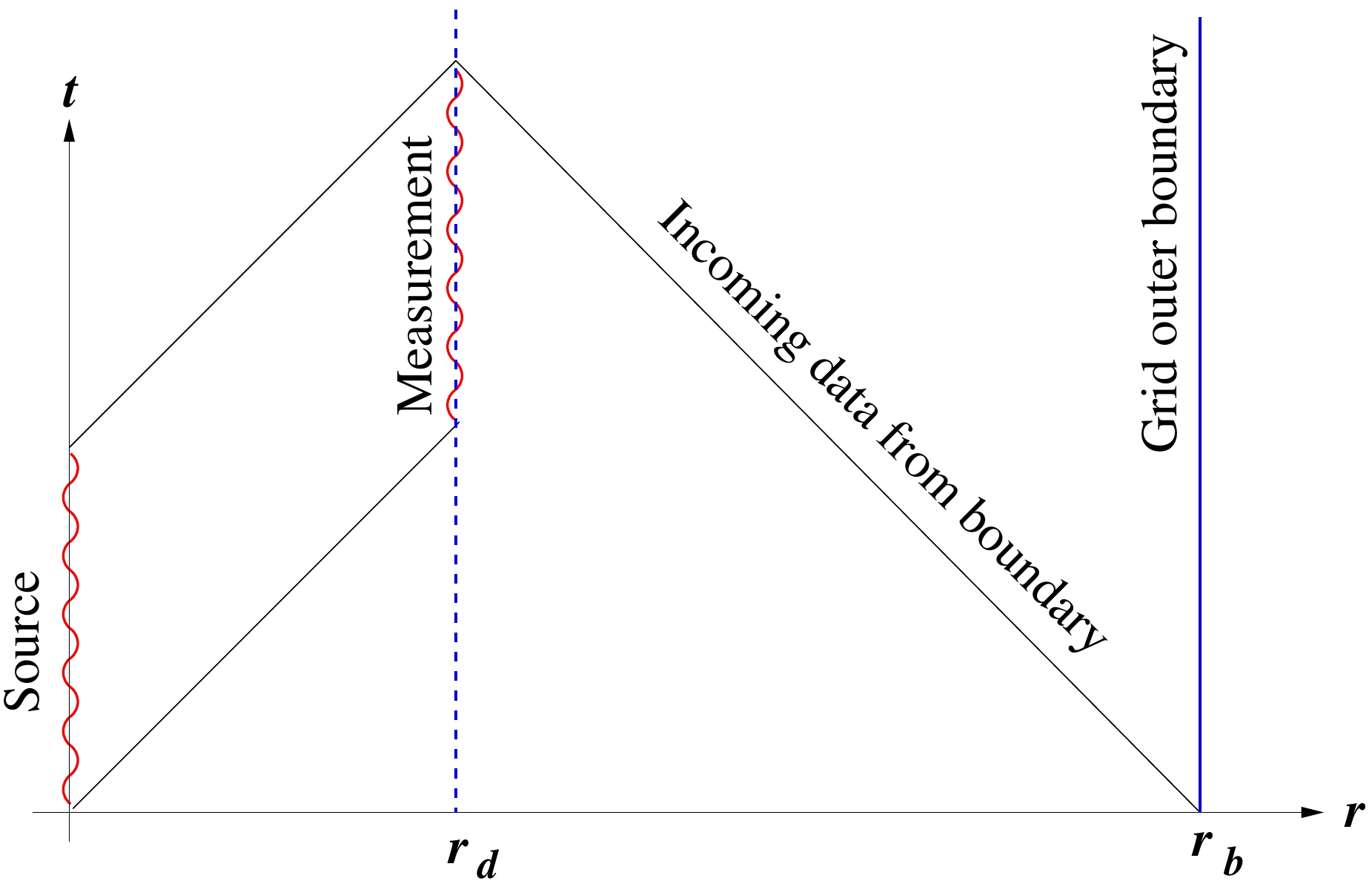}
  \end{center}
  \caption{Schematic of the causal propagation of information during
    the evolution. The gravitational wave source is located in the
    vicinity of $r=0$, with waves propagating outward at the speed of
    light $c=1$, and are measured at radius $r_d$ for a time of
    interest which would include the inspiral, merger and ringdown of
    a binary system. The unphysical outer boundary of the grid is
    located at $r_b$, which is chosen to be sufficiently far removed
    that the future Cauchy horizon of the domain of dependence of the
    initial slice does not reach $r_b$ until the measurement is complete.}
  \label{fig:causal_bc}
\end{figure}

The outer boundary for the spherical grids was chosen based on the
expected time duration of the measurement and radius of the furthest
detector, in order to remove any influence of the artificial outer
boundary condition. In particular, given that the evolution takes a
time $T_m$ for the entire inspiral, merger and ringdown, and
gravitational wave measurements taken at a finite radius $r_d$, we
would like to ensure that a disturbance travelling at the speed of
light from the outer boundary does not reach the measurement radius
(see Fig.~\ref{fig:causal_bc}). That is, noting that the physical
modes travel at the speed of light, $c=1$~\cite{Brown2007a,
  Brown2007b}\footnote{The $1+\log$ slicing condition which we use
  propagates at $\sqrt{2}c$~\cite{Bona95b}, however this is a gauge
  mode and empirically we find it to have negligible effect on
  measurements.}, we place the boundary at
\begin{equation}
  r_b > T_m + 2 r_d.
\end{equation}
For the particular evolution considered here, $T_m\simeq 1350M$, and
our outermost measurements are taken at $r_d=1000M$. We have placed
the outer boundary of the evolution domain at $r_b=3600M$.

The near-zone grids incorporate 5 levels of 2:1 mesh refinement,
covering regions centred around each of the black holes. For the
highest resolution we have considered here, the finest grid (covering
the black hole horizon) has a grid spacing of $dx=0.02M$.  The wave-zone grids
have an inner radial resolution which is commensurate with the coarse
Cartesian grid resolution, $dr=0.64M$ in this case. This resolution is
maintained essentially constant to the outermost measurement radius
($r=1000M$), at which point we apply a gradual decrease in resolution
(as described in Sec.~\ref{sec:coords}) over a distance of $r=500M$.
From $r=1500M$ to the outer boundary, we maintain a resolution of
$dx=2.56M$, sufficient to resolve the inspiral frequencies of the
dominant $(\ell,m)=(2,2)$ mode of the gravitational wave signal. The
transition between the resolutions is performed over a distance of
$500M$ between $r=1000M$ and $r=1500M$. The
angular coordinates have 31 points (30 cells) in $\nu$ and $\phi$ on
each of the 6 patches. The time-step of the wave-zone grids is
$dt=0.144$, and we take wave measurements at each iteration.

We have carried out evolutions at three resolutions in order to
estimate the convergence of our numerical methods. The grid described
above is labelled $h_{0.64}$. The lower resolutions, labelled
$h_{0.80}$ and $h_{0.96}$ have each of the specified grid spacings
scaled by $0.80/0.64$ and $0.96/0.64$, respectively.

\subsection{Results}

\begin{figure*}
  \begin{center}
    \includegraphics[width=86mm,clip,trim=10 0 0 0]{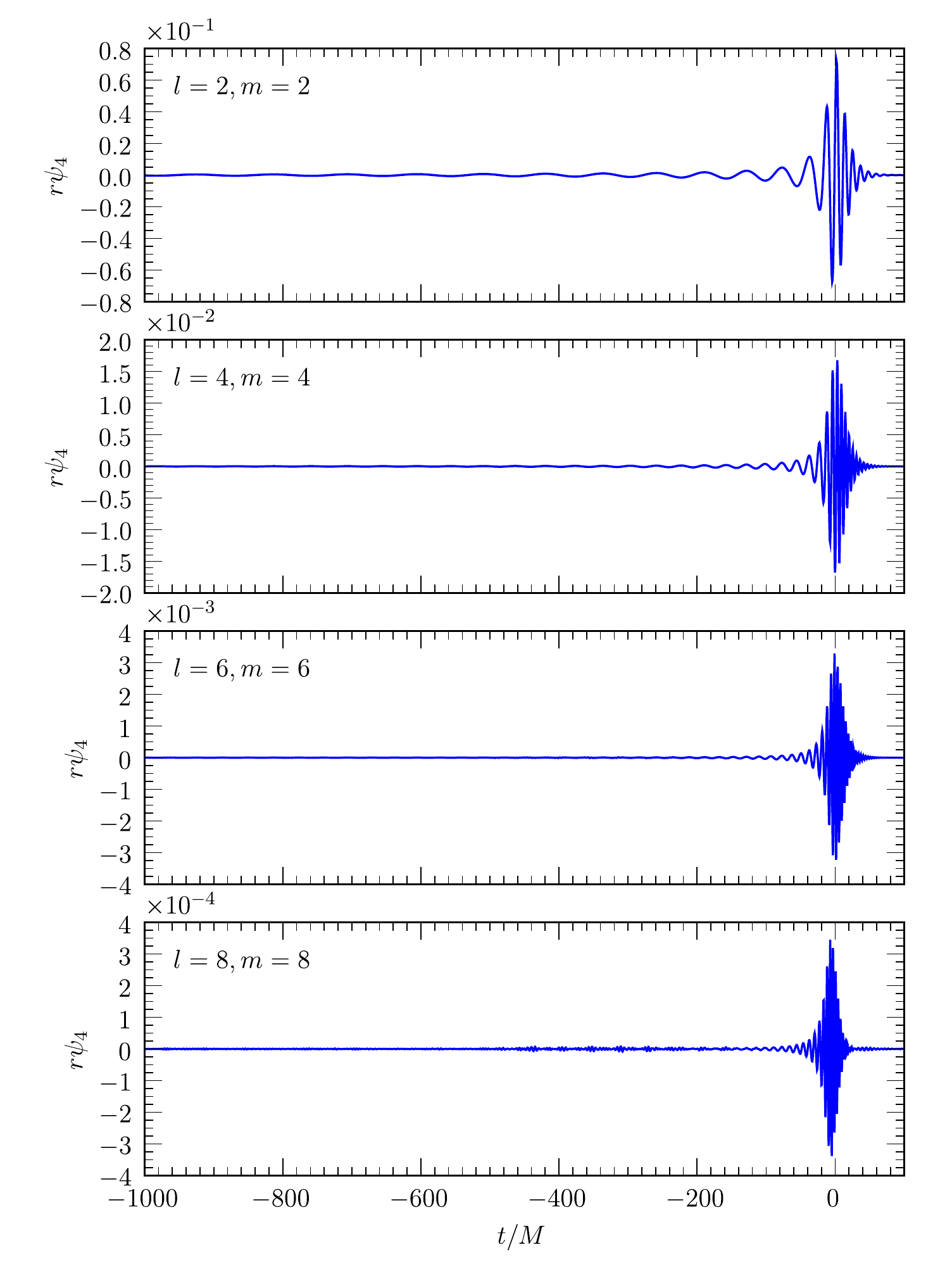}
    \includegraphics[width=86mm,clip,trim=5 0 0 0]{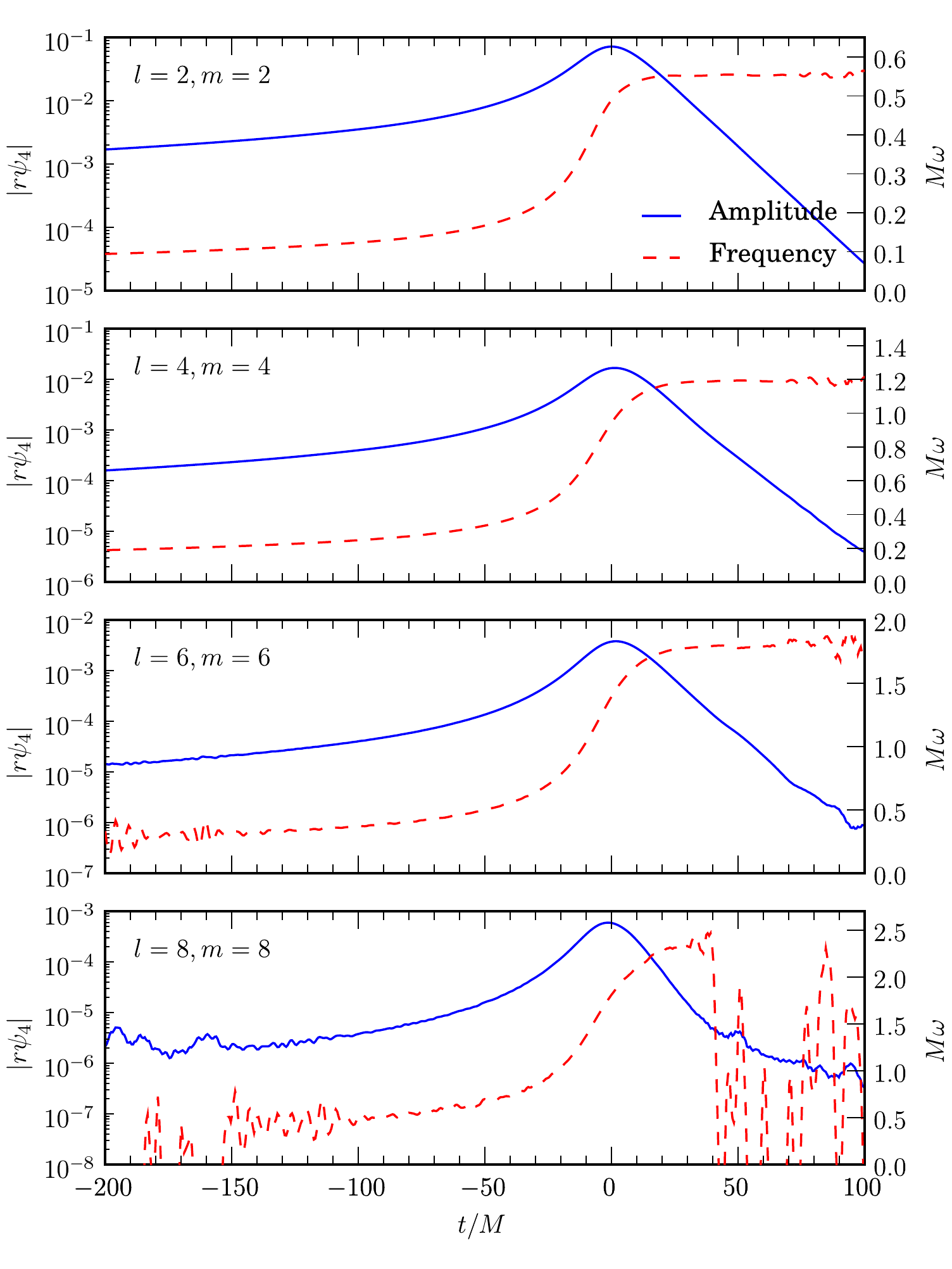}
  \end{center}
  \vspace{-0.5cm}
  \caption{The dominant spherical harmonic modes of $\psi_4$ for
    $\ell=2,4,6,8$, measured at $r=200M$ from the coordinate
    centre. The plots on the right show amplitude and frequency
    evolution during the late inspiral, merger and ringdown..}
  \label{fig:psi4-modes}
\end{figure*}

The binary black hole initial data described in
Sec.~\ref{sec:initial_data} evolves for about 8 orbits ($\simeq
1350M$) before merger. Various $(\ell,m)$ modes of $\psi_4$ are
plotted in Fig.~\ref{fig:psi4-modes}. We find that for the grids we
have used, the modes to $(\ell,m)=(4,4)$ mode are quite well resolved
throughout the evolution. The $(6,6)$ mode is also measurable, and
shows a clear signal, particularly during ringdown.  The $(8,8)$ mode
is dominated by noise for most of the inspiral, though during the
merger and ringdown phase, a clear signal is present and the amplitude
and frequency can be estimated. Tests with an analytical solution
confirm that the angular resolutions which we have used are at best
marginal for resolving this mode.

In the following sections, we report results regarding the convergence
and accuracy of these measurements, as well as determine the
parameters of the merger remnant. By analysing the ring-down behaviour
of the waves we conclude that the remnant is indeed a Kerr black hole
(see Sec.~\ref{sec:ringdown}, below).

\subsubsection{Numerical convergence}

We can establish the consistency of our discretisation by showing that
it does indeed converge to a unique solution in the continuum limit.
Ideally, an exact solution can be used to test this. However, since
there are no exact solutions which adequately model the physical
scenario which we wish to consider (inspiralling black hole binaries),
an alternative is to evaluate numerical solutions at several (at least
three) different resolutions and establish that the differences
decrease as resolution is increased.  For an implementation in which
all of the discrete operations are carried out with the same order of
accuracy, the convergence test should yield a clear exponent
corresponding to that order.

The evolution code incorporates a number of discrete operations, which
for various practical reasons, are carried out to different orders of
accuracy. These are listed in Table~\ref{tab:conv-order}. The primary
operation which is carried out over the bulk of the grid is the
computation of finite difference derivative operations in order to
evaluate the right-hand side of the evolution
equations~\eqref{eq:evo_phi}--\eqref{eq:evo_tG}. For the tests carried
out in this paper, 8th-order stencils are used for this operation,
including the upwinded advection terms.  It is common to apply a small
amount of artificial dissipation in order to smooth high-frequency
effects. This is done at one higher order (9th) than the interior
finite differencing in order to maintain the correct continuum limit.
(In our experiments, however, we have noted that dissipation at this
high order has a negligible impact on the solution, and can
effectively be omitted.) Various boundary operations (inter-patch
boundary communication, mesh-refinement boundaries) are carried out at
lower order. This is done largely for efficiency reasons, as the
communication involved in boundary interpolation can be time-consuming
if the stencil widths are large. Intuitively, the numerical error
associated with these operations may have reduced influence in any
case, as they are applied only at 2D interfaces. In practise this does
seem to be the case, for instance, as experiments with 4th and 5th
order interpolation operators between patches show similar accuracy in
the final solution. Similarly, operations involving different
time-levels are at lower order, again for efficiency reasons. The time
resolution of our evolutions tends to be high enough that one might
expect a small error coefficient of the RK4 integrator. The lowest
order operation which we use is the 2nd-order time interpolation at
mesh-refinement boundaries. Applying higher order here would require
keeping more time levels in memory (currently we store three). Our
results are consistent with previous studies using mesh-refinement for
black hole evolution which suggest that the influence of the low order
time-interpolation boundary conditions is negligible for the time
resolutions which we apply (see, for example, \cite{Brugmann:2008zz}).

\begin{table}
  \begin{ruledtabular}
    \begin{tabular}{lc}
      Numerical method                  & Order \\ \hline
      Grid interior finite differencing & 8 \\
      Inter-patch interpolation          & 5 \\
      Kreiss-Oliger dissipation         & 9 \\
      Time integration (RK4)            & 4 \\
      Mesh-refinement:                  &   \\
      \hspace{5mm} Spatial prolongation & 5 \\
      \hspace{5mm} Spatial restriction  & n/a \\
      \hspace{5mm} Time interpolation   & 2 \\
      Analysis tools:                   &   \\
      \hspace{5mm} Interpolation        & 4 \\
      \hspace{5mm} Finite differencing  & 8 \\
      \hspace{5mm} Surface integration  & $2N-1$ \\
    \end{tabular}
  \end{ruledtabular}
  \caption{\label{tab:conv-order}Table of convergence order of various
    numerical aspects of the evolution code. Spatial restriction is
    carried out by a direct copy. The surface integration is exact for
    polynomials up to degree $2N-1$, where $N$ is the number of
    grid-points along one direction on the sphere.}
\end{table}

For test cases involving a single non-spinning black hole, in fact we
find 8th-order convergence in the Hamiltonian constraints. This is
likely due to the relatively constant values (except for some gauge
evolution) maintained by the evolution variables during the evolution,
which minimises error due to time-integration or propagation across
boundaries.

\begin{figure*}
  \begin{center}
    \includegraphics[width=190mm,clip,trim=100 0 50 0]
      {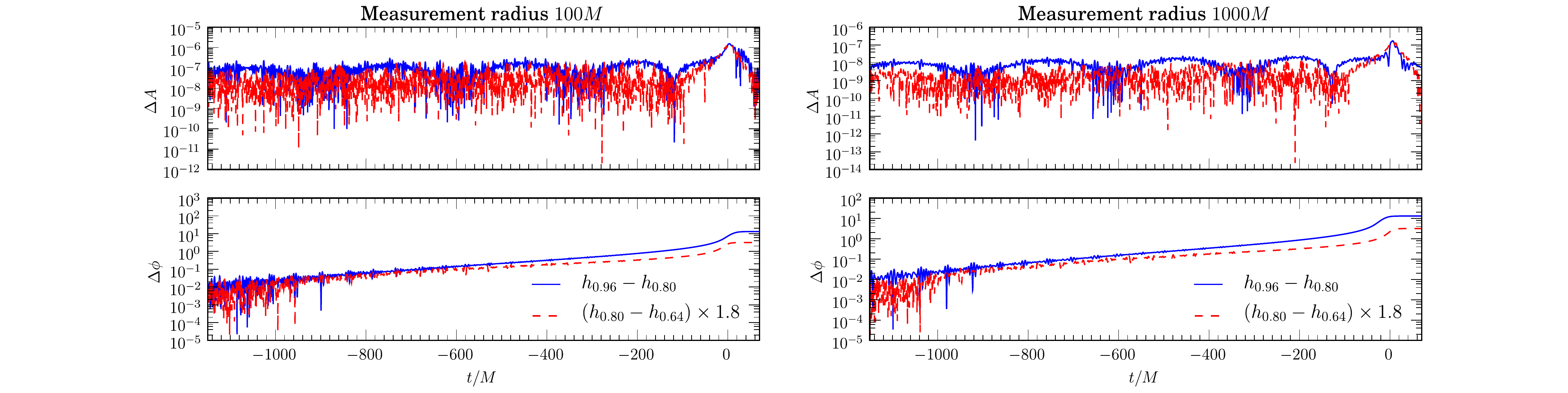}
  \end{center}
  \vspace{-\baselineskip}
  \caption{Convergence in amplitude (top) and phase (bottom) of the
    $(\ell,m)=(2,2)$ mode of $\psi_4$ for detectors at $r=100M$ and
    $r=1000M$. The higher resolution difference, $h_{0.80}-h_{0.64}$, is
    scaled for 4th-order convergence.}
  \label{fig:d550_convergence_22}
\end{figure*}

A more relevant situation is that of a binary black hole inspiral,
which we have tested using the parameters described above in
Sec.~\ref{sec:initial_data}. For this model, we have measured the
gravitational waveform, $\psi_4$, integrated over spheres at radii
from $r=100M$ to $r=1000M$, at the three resolutions $h_{0.96}$,
$h_{0.80}$ and $h_{0.64}$. Results for the $(\ell,m)=(2,2)$ mode are shown in
Fig.~\ref{fig:d550_convergence_22}. The evolution lasts for about
$1350M$ before merger, and the plots encompass the inspiral, merger
(at $t=0M$ on this time axis), and ringdown. The figure plots the
error in phase $\Delta\phi$ and relative amplitude $\Delta A$ for
the $(2,2)$ mode extracted at $r=100M$ and $r=1000M$,
respectively, between medium $h_{0.80}$ and low $h_{0.96}$ resolutions and
high $h_{0.64}$ and medium $h_{0.80}$ resolutions in the wave-zone.  The
latter error is scaled such that the curves will overlap in the
case of a 4th-order convergent solution. At both radii, we
find that during the inspiral phase, the rescaled error of the higher
resolutions lies below that of the lower resolution, suggesting better
than 4th-order convergence (in fact, closer to 8th-order over significant
portions of the plot). At later times, around the peak of the
waveform, the curves are more closely aligned, indicating 4th-order
convergence. The plot suggests that during the very dynamical late
stages of the inspiral, the lower order boundary conditions and/or 
the time integration, play a more important role relative to the early
inspiral phase of the evolution, where the convergence order is
closer to that of the interior finite differencing. The results are,
however, convergent over the entire evolution (including merger and
ringdown). As we will see in the next section, the accuracy is
excellent for these resolutions so that the rate of convergence
is not a particular issue.

\begin{figure}
  \begin{center}
    \includegraphics[width=1.\linewidth,clip,trim=25 0 25 0]
      {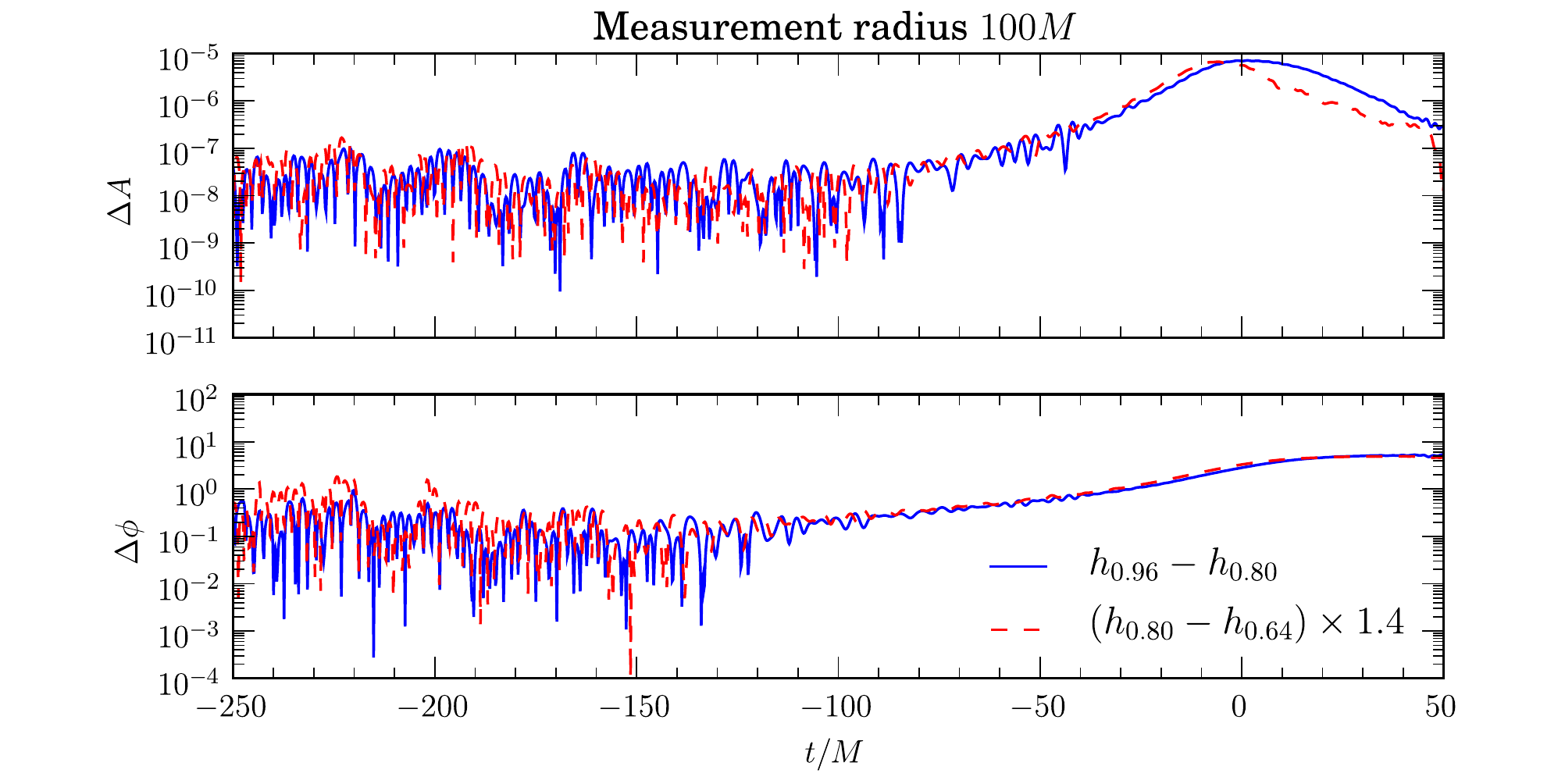}
  \end{center}
  \vspace{-\baselineskip}
  \caption{Convergence in amplitude (top) and phase (bottom) of the
    $(\ell,m)=(6,6)$ mode of $\psi_4$ for detector at $r=100M$ during
    the late through merger. The higher resolution difference,
    $h_{0.80}-h_{0.64}$, is scaled for 3rd-order convergence.}
  \label{fig:d550_convergence_66}
\end{figure}

We have verified convergence for a number of different modes of the
$\psi_4$ waveform at different radii. For instance,
Fig.~\ref{fig:d550_convergence_66} shows similar results for the
$(\ell,m)=(6,6)$ mode, which is some two orders of magnitude smaller
in peak amplitude than the $(\ell,m)=(2,2)$ mode (see
Fig.~\ref{fig:psi4-modes}). During the early inspiral, it is difficult
to evaluate a convergence order due to high frequency noise which
is large relative to the waveform amplitude. However, a measurable
signal is clear in the last orbit, merger and ringdown phase, and
converges at a clear 3rd order.

\subsubsection{Accuracy}
\label{sec:accuracy}

We estimate the numerical phase and amplitude error by means of
a Richardson expansion at a given resolution $\Delta$, 
\begin{equation}
u_\Delta(t,x)=u(t,x)+\Delta e_1(t,x)+\Delta^2 e_2(t,x)+\cdot\cdot\cdot\,,
\label{eq:Richardson-expand}
\end{equation}
where $u(t,x)$ is the solution of the original differential equation,
and the $e_i(t,x)$ are error terms at different orders in $\Delta$.
Assuming convergence at a fixed order, $n$, we can expect some of
these error functions to vanish. Using solutions, $u$, obtained at
two resolutions, $\Delta_1$ and $\Delta_2$, the Richardson expansion
implies
\begin{align}
  u_{\Delta_1}-u_{\Delta_2} 
  & = e_n(\Delta_1^n-\Delta_2^n) + \mathcal{O}(\Delta^{n+1}) \nonumber \\
  & = e_n\Delta_2^n(C^n-1) + \mathcal{O}(\Delta^{n+1}) \nonumber \\
  & \sim \epsilon_{\Delta_2}(C^n-1)\,,
\end{align}
where $\epsilon_{\Delta_2}$ is the estimated solution error on the
higher resolution grid, and where
\begin{equation}
  C^n := \left(\frac{\Delta_1}{\Delta_2}\right)^n\,.
\end{equation}

We thus obtain an estimate for the solution error that is at least
accurate to order $n+1$,
\begin{equation}
  \epsilon_{\Delta_2}\sim\frac{1}{C^n-1}(u_{\Delta_1}-u_{\Delta_2})\,,
  \label{eq:numerical-error}
\end{equation}
which we use as an estimate of the numerical error in our solutions.

\begin{figure}
  \begin{center}
    \includegraphics[width=1.\linewidth,clip,trim=25 0 25 0]
      {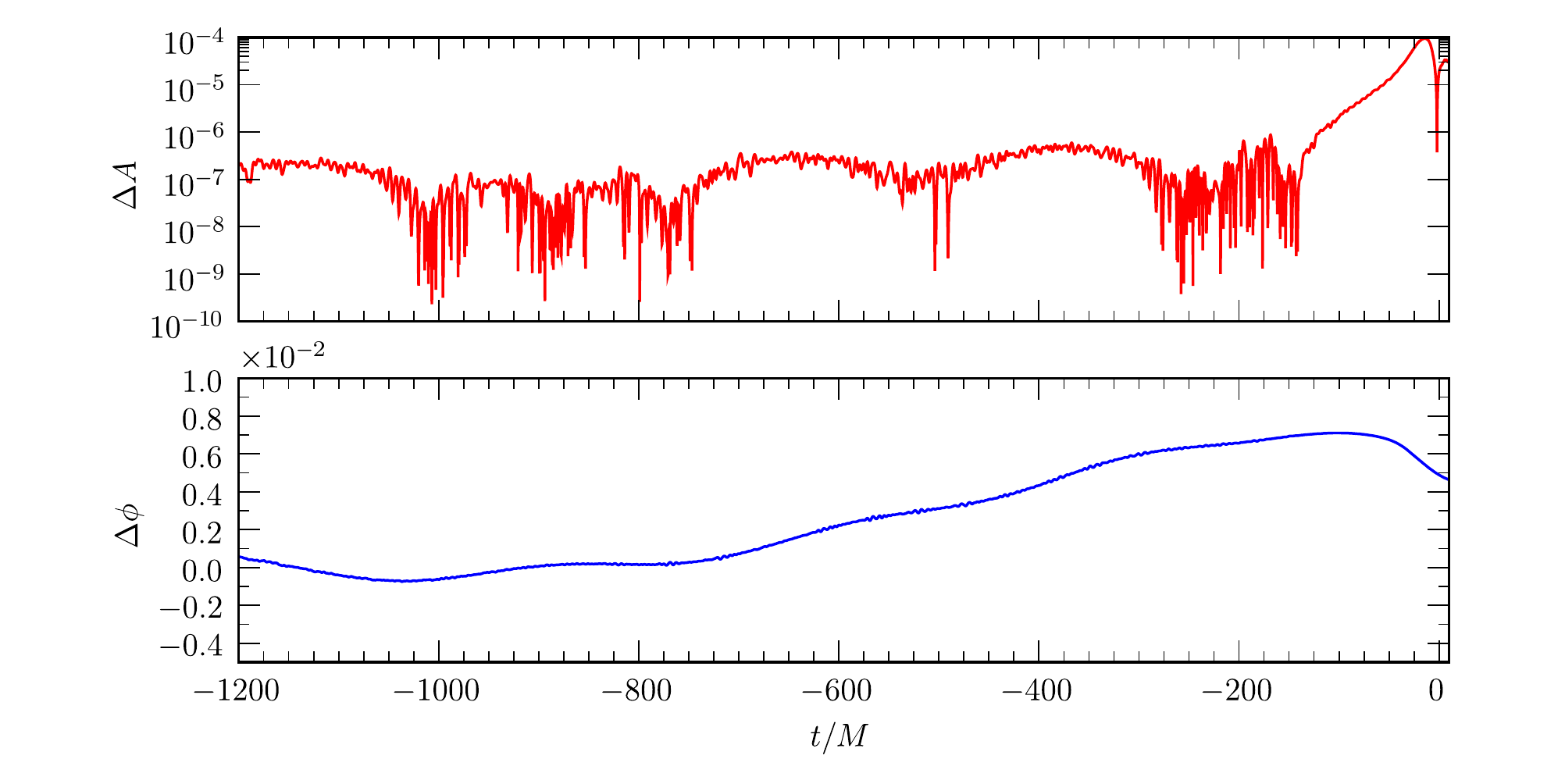}
  \end{center}
  \vspace{-\baselineskip}
  \caption{Absolute numerical error in the amplitude (top) and
    phase (bottom) accumulated over the course of the evolution for
    the highest resolution run, determined according to
    Eq.~\eqref{eq:numerical-error} for the point-wise differences in
    amplitude and phase between medium and high resolution runs. For
    the phase we assume the measured 8th-order convergence over the
    entire evolution, while for the amplitude we use 8th-order before
    $t\leq-100$, and 4th-order thereafter (see text).}
  \label{fig:error}
\end{figure}

During the inspiral phase (which for this purpose we regard as being
the period $t\leq-100M$), we have found roughly 8th-order convergence
in the amplitude and phase, as described above. The remaining relative
error for the $(\ell,m)=(2,2)$ mode can be estimated as
\begin{subequations}
\begin{align}
  \max_{T\in[-1350,-100]}{\text{err}(A)}_{\rm inspiral}    & = 0.090\%\,, \\
  \max_{T\in[-1350,-100]}{\text{err}(\phi)}_{\rm inspiral} & = 0.010\%\,.  %0.005 \rad\,,
\end{align}
\end{subequations}
where $\text{err}(A) := \Delta A/A$ and $\text{err}(\phi) :=
\Delta\phi/\phi$, i.e., the rate of loss of phase with $\phi$.
During merger and ring-down ($t>-100M$), we observe 4th-order
convergence in the amplitude, while maintaining 8th-order convergence
in the phase. This results in the estimate
\begin{subequations}
\begin{align}
  \max_{T\in(-100,150]}{\text{err}(A)}_{\rm merger} & = 0.153\%\,, \\
  \max_{T\in(-100,150]}{\text{err}(\phi)}_{\rm merger} & = 0.003\%\,.  %0.002 \rad\,.
\end{align}
\end{subequations}
The time evolution of the numerical error in phase and amplitude is
shown in Fig.~\ref{fig:error}.

We note that these errors are of comparable order to the errors
inherent in the extrapolation~\cite{Pollney:2009MP-unpublished}.
Moreover, as is pointed out in~\cite{Reisswig:2009us}, the error
between extrapolated waveforms and those determined at future null
infinity, $\scri$, by characteristic extraction, is an order of
magnitude larger than the numerical error determined here. This
highlights the importance of reducing systematic errors inherent
in finite radius measurements of $\psi_4$.

\subsubsection{Properties of the merger remnant}

The merger remnant can be measured with high accuracy, using either
the isolated horizon formalism~\cite{Dreyer02a, Ashtekar:2004cn}, or
geometrical measures of the apparent horizon~\cite{Brandt94c,
  Alcubierre:2004hr}. Some results are reported in
Table~\ref{tab:remnant}, along with estimated numerical errors. The
results agree well with previous high-accuracy measurements, such as
those obtained by spectral evolution~\cite{Scheel:2008rj,
  Hannam:2009hh}, with the spin and irreducible mass agreeing within
three decimal and four decimal places, respectively.  places. While
this is larger than the reported errors, we note that we have evolved
a different initial data set than~\cite{Scheel:2008rj}. As reported in
Sec.~\ref{sec:initial_data} our evolution has somewhat more
eccentricity, and the level of agreement can be used to judge the
influence of small amounts of eccentricity on the result.

By comparing the properties of the merger remnant with the integrated
radiated energy, $E_{\rm rad}$, and angular momentum, $J_{\rm rad}$,
determined from the gravitational waveforms, we find the residuals
\begin{subequations}
  \begin{align}
    |M_f + M_{\rm rad} - M_{\rm ADM}| & = 2.6 \times 10^{-4}, \\
    |S_f + J_{\rm rad} - J_{\rm ADM}| & = 3.1 \times 10^{-4}.
  \end{align}
\end{subequations}
Here we have used the extrapolations of the gravitational waveforms to
$r\rightarrow\infty$ based on the 6 outermost measurement radii. A
more detailed discussion of this procedure is given
in~\cite{Pollney:2009MP-unpublished}. The results can be further
improved through taking measurements at $\scri$, as outlined
in~\cite{Reisswig:2009us, Reisswig:2009-cce-long}.

\begin{table}
  \begin{ruledtabular}
    \begin{tabular}{lD{.}{.}{2.20}}
      Total ADM mass, $M_{\rm ADM}$  &  0.99051968 \pm 20\times 10^{-9} \\
      Total ADM angular momentum, $J_{\rm ADM}$
                                     &  0.99330000 \pm 10\times 10^{-17} \\ \hline
      Irreducible mass, $M_{\rm irr}$&  0.884355 \pm 20\times 10^{-6} \\
      Spin, $S_f/M_f^2$              &  0.686923 \pm 10\times 10^{-6} \\
      Christodoulou mass, $M_{\rm f}$&  0.951764 \pm 20\times 10^{-6} \\
      Angular momentum, $S_f$        &  0.622252 \pm 10\times 10^{-6} \\ \hline
      Radiated energy, $E_{\rm rad}$ &  0.038546 \pm 51\times 10^{-6} \\
      Radiated angular momentum, $J_{\rm rad}$
                                     &  0.370391 \pm 17\times 10^{-6} \\
    \end{tabular}
  \end{ruledtabular}
  \caption{Properties of the merger remnant as measured on the
    apparent horizon ($M_{\rm irr}$, $S_f/M_f^2$) and from the
    gravitational radiation ($E_{\rm rad}$, $J_{\rm rad}$). Ranges
    indicate the estimated numerical error. For the error in
    $J_{\rm ADM}$, we have simply quoted machine precision
    (it is an analytical expression of the input momenta on the
    conformally flat initial slice).}
  \label{tab:remnant}
\end{table}

\subsubsection{Quasi-normal modes of the merger remnant}
\label{sec:ringdown}

In Fig.~\ref{fig:psi4-modes}, we have shown the late-time
behaviour of the amplitude and frequency for the dominant spherical
harmonic modes of $\psi_4$, to $(\ell,m)=(8,8)$.  We note that during
ring-down, the frequencies settle to a constant value. If the final
black hole is a Kerr black hole, these frequencies are given by the
quasi-normal modes of a Kerr black hole with given spin $a$.

As reported in the previous section, our evolution leads to a merger
remnant with $a=0.686923\pm 1\times 10^{-5}$ (see
Table~\ref{tab:remnant}), as measured on the horizon.  The real part
of the prograde quasi-normal mode (QNM) frequencies for modes up to
$(\ell,m)=(7,7)$, can be found tabulated in~\cite{Berti06c}.
For example, $M\omega_{22}=0.526891$ for the $(\ell,m)=(2,2)$ mode,
given a final black hole of the measured mass $M_f$ and spin $S_f$.

At this point it is worth noting that the QNM determined from
perturbations of a Kerr black hole are most naturally expressed in
terms of a basis of spin-weighted \textit{spheroidal} harmonics.  By
contrast, our waveforms have been decomposed relative to a basis of
spin-weighted \textit{spherical} harmonics, which are easily
calculated via Legendre functions. In order to make an appropriate
comparison between these modes with the perturbative results
we need to apply a transformation to the wave-modes. We have
\begin{equation}
  \hat{\psi}_4^{\ell'm'} 
  = \sum_{\ell,m} \psi_4^{\ell,m} \langle \ell,m| \ell',m' \rangle\,,
\end{equation}
where a dash denotes labelling of the spheroidal harmonic modes, and
$\langle \ell,m| \ell',m' \rangle$ is the overlap defined by
\begin{equation}
  \langle \ell,m| \ell',m' \rangle
  = \int_\Omega d\Omega {}_{-2}\bar{S}_{\ell' m'}(c_{\ell'm'}) {}_{-2}Y_{\ell m}\,.
\end{equation}
The spheroidal harmonics parameter $c_{\ell'm'}=a \omega_{\ell' m'}$
depends on the spin $a$ of the black hole and the corresponding
prograde or retrograde QNM frequency $\omega_{\ell' m'}$ of the
$(\ell' m')$ spheroidal harmonic mode\footnote{We restrict attention
  to the $N=0$ harmonic only.}. If $c=0$ (as is the case for
non-spinning black holes), the spheroidal harmonics reduce to the
spherical harmonics.  The spin-weighted spheroidal harmonics used here
have been implemented following Leaver~\cite{Leaver85} and
are reviewed in \cite{Berti06c}.

\begin{figure}
  \begin{center}
    \includegraphics[width=86mm,clip,trim=5 50 0 40] %5 50 36 50]
      {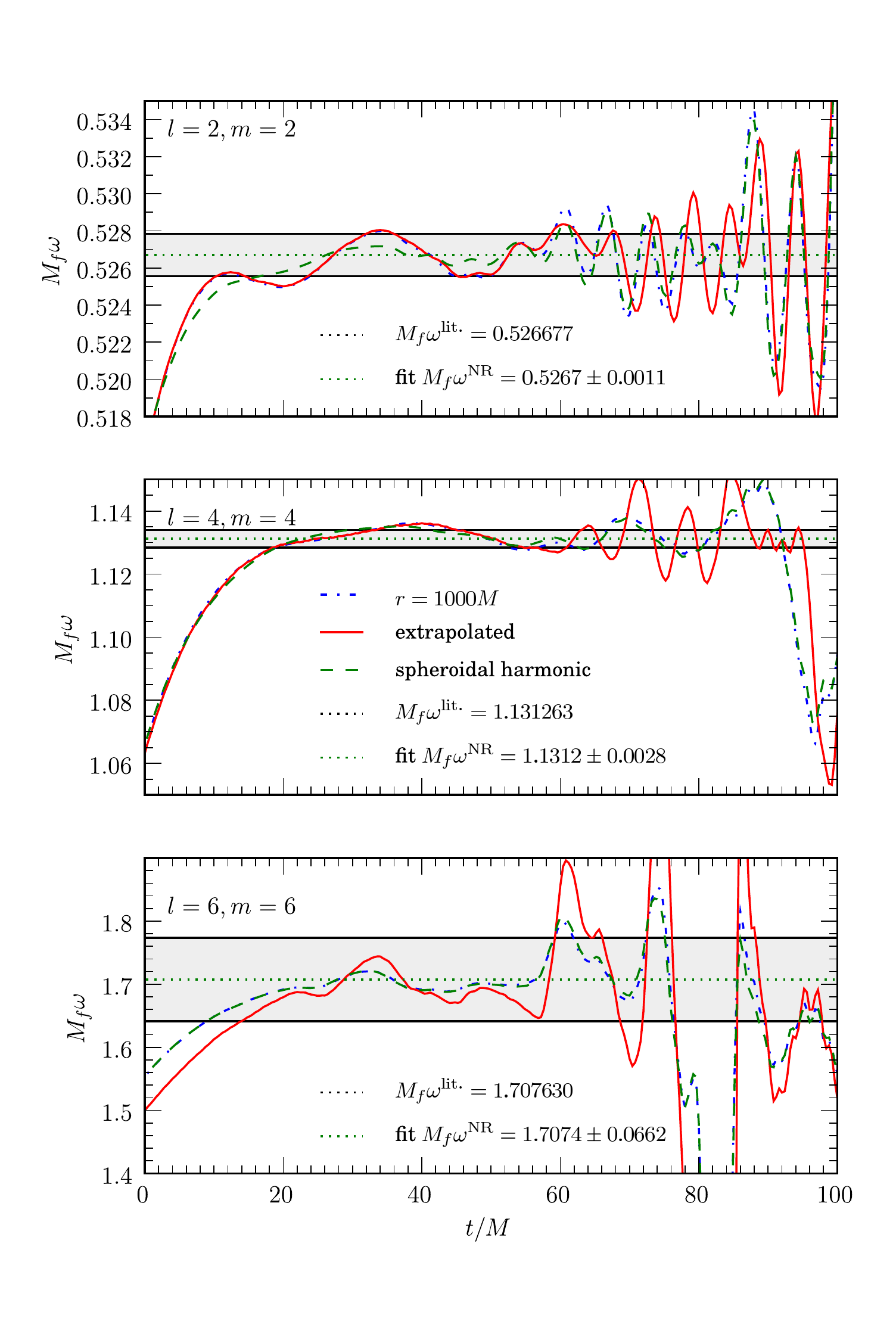}
  \end{center}
  \vspace{-\baselineskip}
  \caption{The ringdown frequencies for the dominant $\psi_4$ modes to
    $\ell=6$ of the merger remnant. From top to bottom, the plots show
    the frequencies of the $(\ell,m)=(2,2)$, $(4,4)$ and $(6,6)$ modes
    respectively, over a timescale from the $(2,2)$ waveform peak to
    $100M$ later, at which point the waveform amplitude is too small
    to measure an accurate frequency. The $\psi_4$ data measured at
    $r=1000M$ is plotted, in addition to the value extrapolated to
    $r\rightarrow\infty$, and the transformation to spheroidal
    harmonics. The expected quasi-normal mode frequency is plotted as
    a dotted line, as well as a fit to the spheroidal harmonic data
    over the range $t\in[40M,80M]$, with error-bars determined by the
    standard deviation of the fit.}
  \label{fig:psi4_amp_freq_modes-ringdown}
\end{figure}

The frequencies measured during the ringdown are plotted in
Fig.~\ref{fig:psi4_amp_freq_modes-ringdown} for the modes
$(\ell,m)=(2,2)$,$(4,4)$ and $(6,6)$. We have plotted data for the
$r=1000M$ measurement, as well as the value obtained by extrapolating
the waveforms extracted at the outermost 6 measurement spheres to 
$r\rightarrow\infty$, and find that in fact the extrapolation has little
effect on the frequency of the lower order modes at these distances from the
source. We note that there is a modulation of the ringdown frequency,
particularly apparent in the $(2,2)$ mode. This is a result of mode
mixing, which stems from the use of the spherical harmonic basis for
the $\psi_4$ measurements. By transforming the $r=1000M$ result to
spheroidal harmonics, this modulation visible in the $t<40M$ signal
is largely removed (dashed line).

As the amplitude of the wave declines exponentially to the level of
numerical error, the frequencies become difficult to measure
accurately. We estimate the ringdown frequency for each mode by
performing a least-squares fit of a horizontal line through the
measured spheroidal harmonic frequency over the range $t\in[40,80]M$
(dotted line) with the standard deviation of the fit as a gauge of the
error (grey region). These constant lines represent the estimated
frequency of the associated QNM modes, and are tabulated as
$\omega^{\rm NR}$ in Table~\ref{tbl:ringdown}. They agree to high
precision with the prograde QNM frequencies, $\omega^{\rm lit.}$,
determined Kerr black holes by perturbative methods~\cite{Berti06c}.
We conclude that the merger remnant is compatible with a Kerr black
hole within the given error estimates.

\begin{table}
  \begin{ruledtabular}
    \begin{tabular}{lccc}
      $(\ell,m)$ &  $M_f\omega^{\rm lit.}$ & $M_f\omega^{\rm NR}$   & $|M_f\omega^{\rm NR}-M_f\omega^{\rm lit.}|$  \\ \hline
      $(2,2)$    &  $0.526891$       & $0.5267 \pm 0.0011$ & $1.9\times10^{-4}$ \\
      $(4,4)$    &  $1.131263$       & $1.1312 \pm 0.0028$ & $6.3\times10^{-5}$ \\
      $(6,6)$    &  $1.707630$       & $1.7074 \pm 0.0662$ & $2.3\times10^{-4}$
    \end{tabular}
  \end{ruledtabular}
  \caption{\label{tab:qnms}Prograde $N=0$ QNM frequencies for
    different modes and spin $a=0.6869$ as determined by
    perturbative methods~\cite{Berti06c}, $\omega^{\rm lit.}$, and as
    measured during ringdown in the numerical relativity simulation,
    $\omega^{\rm NR}$.}
  \label{tbl:ringdown}
\end{table}

%%%%%%%%%%%%%%%%%%%%%%%%%%%%%%%%%%%%%%%%%%%%%%%%%%%%%%%%%%%%%%%%%%%%%%%%%%%%%%
\section{Discussion}
\label{sec:discussion}

The results of this paper provide a demonstration of the usefulness of
adapted coordinates in numerical relativity simulations. The precision
of the calculations have allowed us to obtain convergent modes 
to $\ell=6$, through merger and ringdown, with accurate predictions
of the quasi-normal ringdown frequencies of the remnant.

Our implementation of non-singular radially adapted coordinates for
the wave zone is based on the use of multiple grid patches with
interpolating boundaries, coupled to a BSSNOK evolution
code. Thornburg~\cite{Thornburg:2004dv} first demonstrated that such a
setup could lead to stable evolutions in the case of a spinning black
hole in Kerr-Schild coordinates. We have demonstrated that the
approach is also effective and robust for dynamical puncture
evolutions, and in particular the problem of binary black holes.

The implementation described here has a number of advantages,
principle among them being its flexibility. While we have presented
results for a particular grid structure adapted to radially
propagating waves, there are no principle problems with restructuring
the grids to cover any required domain, for instance adapted to
excision boundaries or toroidal fields. Since data is stored in the
underlying Cartesian basis, and passed by interpolation across
boundaries, the coordinates used on each patch are largely independent
of the others, and there is no need for numerical grid generating
schemes. While we have used the BSSNOK formalism to evolve the
Einstein equations, in principle any stable strongly hyperbolic system
can be substituted. The BSSNOK system has, however, proven particularly
useful for evolving black holes via the puncture approach, which
itself has proven to be a very flexible methodology. We have
demonstrated results for the most well-studied test case, non-spinning
equal-mass black holes, the same techniques can be applied to
different mass ratios and spinning black holes, simply by changing the
physical input parameters. (The appendices include some examples of
spinning black hole evolutions.)

Finally, we emphasise again the accuracies which can be attained by
this approach. Our finite difference results show numerical error
estimates which are on par with those achieved using spectral spatial
discretisation~\cite{Scheel:2008rj}. The adapted radial coordinate
allows us to take measurements at radii much larger than have been
used before, as well as obtain accurate measurements of higher $\ell$
modes during merger, which have an amplitude more than two orders of
magnitude smaller than the dominant $(\ell,m)=(2,2)$ mode. One of the
aspects which makes this possible is the fact that we are able to
extend our grids to a distance such that the measurements are included
in the future domain of dependence of the initial data (causally disconnected
outer boundaries), and the waves are reasonably well resolved over
this entire domain so that internal reflections are
minimised. 
Furher, we note that our results are consistent with other
puncture-method calculation in that the results are convergent and can
be consistently extrapolated to $r\rightarrow\infty$ throughout the
entire evolution, including late inspiral and
ringdown~\cite{Pollney:2009MP-unpublished}, where other approaches
have had difficulties.

The absence of artificial boundaries, as well as dissipative regions
in the wave zone, removes an important source of potential error in
solving the Einstein equations as an initial-boundary value
problem. The remaining errors can be categorised in three
forms. First, numerical error due to the discretisation. This can be
reduced through the use of higher order methods for the operations
performed in various parts of the code, and fortunately is also easy
to quantify by performing tests at multiple resolutions. We note that
for finite differences, the largest improvement in accuracy occurs in
going from 2nd to 4th-order for the interior computations, and beyond
that there are diminishing returns~\cite{Gustafsson95}. While it does
not yet seem to be a limiting factor, except possibly during the
merger, the RK4 time-stepping will at some level of resolution be a
determining factor in the accuracy regardless of the spatial order
(and this is also the case for current implementations of spectral
methods). The second source of error is a physical error, inherent in
the choice of initial data parameters for the binary evolution. At the
separations which are practical for numerical relativity (say
$d<20M$), the physical model is expected to have shed all of its
eccentricity. We have used post-Newtonian orbital parameters to
attempt to place our black holes in low eccentricity trajectories, and
this is quite effective. Alternative approaches, involving iteratively
correcting the initial data parameters until a tolerable eccentricity
has been reached, are able to reduce the eccentricity still
further~\cite{Pfeiffer:2007yz}. This technique can in principle also
be adapted to the moving puncture approach. The final source of error
arises in the measurement of $\psi_4$, which is done at a finite
radius, and then extrapolated to $r\rightarrow\infty$ by some
procedure. We have attempted to minimise this error by placing
detectors at large radii, well into the region where the perturbations
are linear, and have shown that the extrapolations are consistent with
measurements at larger radii, as well as with each other in the
$r\rightarrow\infty$ limit~\cite{Pollney:2009MP-unpublished}. However,
there remain ambiguities particularly in gauge-dependent quantities
such as the choice of surface on which measurements are taken, and the
definition of time and radial distance to be used in the
extrapolation. In a companion paper~\cite{Reisswig:2009us}, we have
demonstrated that these ambiguities can be removed entirely by the
procedure of \emph{characteristic extraction}, whereby evolution data
on a world-tube is used as an inner boundary condition for a fully
relativistic characteristic evolution, extending to null infinity,
$\scri$. The results suggest that systematic errors inherent in finite
radius measurements of $\psi_4$ are more than an order of magnitude
larger than the numerical errors reported here.

%%%%%%%%%%%%%%%%%%%%%%%%%%%%%%%%%%%%%%%%%%%%%%%%%%%%%%%%%%%%%%%%%%%%%%%%%%%%%%

\begin{acknowledgments}
  We dedicate this paper to the memory of Thomas Radke, who has made
  invaluable contributions to the development and optimisation of
  Cactus, Carpet and the code described here.
  The authors are pleased to thank:
    Ian Hinder,
    Sascha Husa,
    Badri Krishnan,
    Philipp Moesta,
    Christian D. Ott,
    Luciano Rezzolla,
    Jennifer Seiler,
    Jonathan Thornburg,
  and Burkhard Zink
  for their helpful input;
  the developers of Cactus \cite{Goodale02a, cactusweb1} and Carpet
  \cite{Schnetter-etal-03b, Schnetter06a, carpetweb} for providing an
  open and optimised computational infrastructure on which we have
  based our code;
  Nico Budewitz for optimisation work with our local compute
  cluster, \texttt{damiana};
  support from the DFG SFB/Transregio~7, the VESF, and by the NSF
  awards no.~0701566 \emph{XiRel} and no.~0721915 \emph{Alpaca}. 
  Computations were performed at the AEI, at LSU,
  on LONI (numrel03), on the TeraGrid (TG-MCA02N014), and the
  Leibniz Rechenzentrum M\"unchen (h0152).
\end{acknowledgments}

%%%%%%%%%%%%%%%%%%%%%%%%%%%%%%%%%%%%%%%%%%%%%%%%%%%%%%%%%%%%%%%%%%%%%%%%%%%%%%
%%% Appendices
%%%%%%%%%%%%%%%%%%%%%%%%%%%%%%%%%%%%%%%%%%%%%%%%%%%%%%%%%%%%%%%%%%%%%%%%%%%%%%
\appendix
\section{The influence of upwinded advection stencils}
\label{sec:upwind}

It has long been recognised that for BSSNOK evolutions employing a
shift vector, $\beta^a$, the overall accuracy can be improved by
``upwinding'' the finite difference stencils for advective terms of
the form $\beta^i\partial_iu$~\cite{Alcubierre02a}. The upwind
derivatives employ stencils which are off-centred by some number of
grid points in the direction of $\beta^a$. The drawback of the method
is that in order to maintain the same order of accuracy in the
derivatives, the stencil must have the same width as a centred
stencil, but since it is offset in either a positive or negative
direction, it effectively requires an additional number of points to
be available to the derivative operator equal to the size of the
offset. For parallel codes which physically decompose the grid over
processors and communicate ghost-zone boundaries, this means that a
larger number of points must be communicated and can impact the
overall efficiency.  Further, a larger number of points must be
translated at inter-patch and refinement level boundaries.

\begin{figure}
  \begin{center}
    \includegraphics[width=\linewidth,clip,trim=00 10 00 30]
      {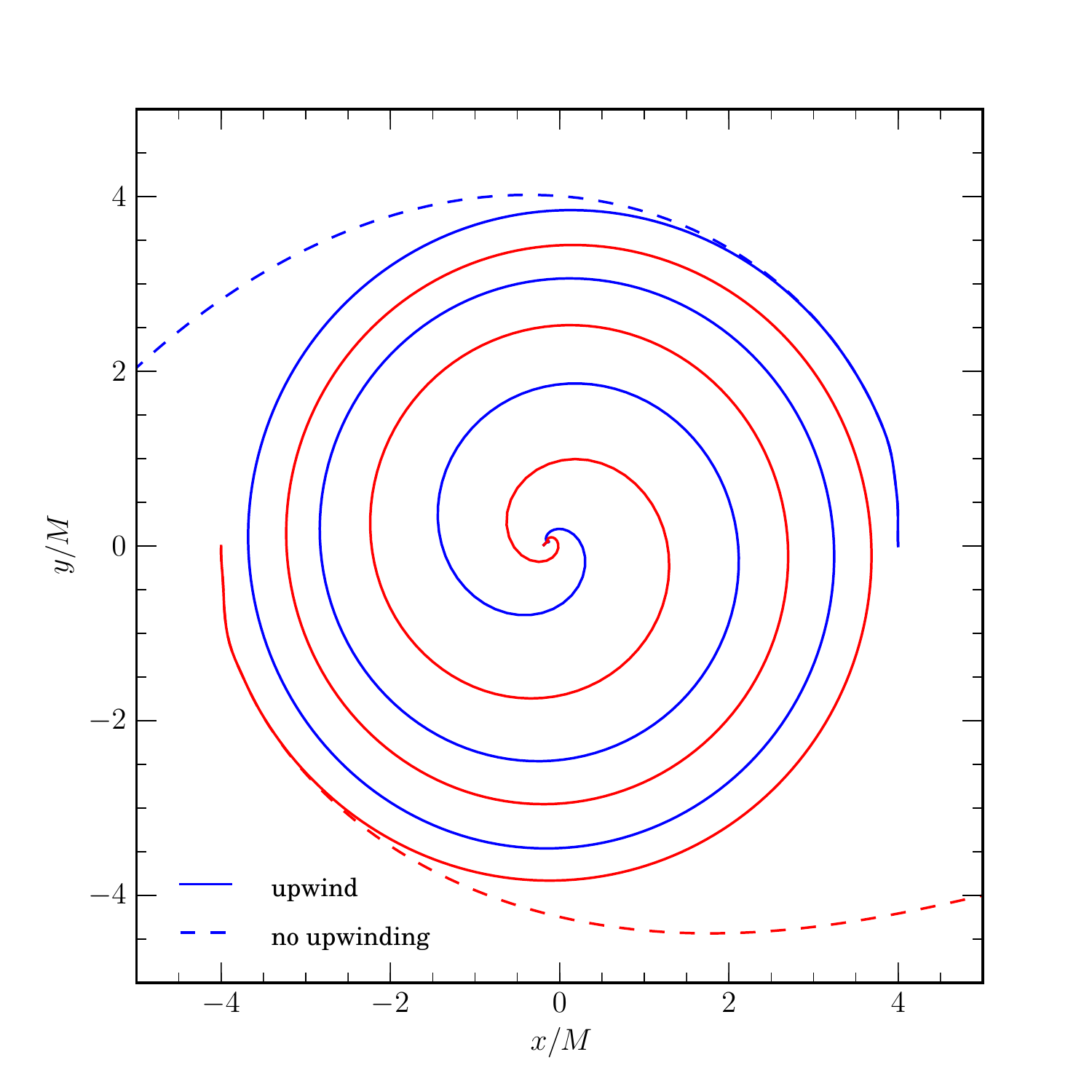}
  \end{center}
  \vspace{-\baselineskip}
  \caption{Trajectories of the two inspiralling punctures for a
    spinning configuration $a_1=-a_2=0.8$, with upwinded advection
    terms (solid lines) and without (dashed lines). In the case where no
    upwinding has been used, the black holes do not inspiral, due to the
    accumulation of numerical error.}
  \label{fig:u8D8-phasing-no-upwinding}
\end{figure}

The original observation that upwinding is helpful was made with a code that
used 2nd-order spatial finite differences. In that case, the centred
stencils are small (three points) and the upwind derivatives
correspond to sideways derivatives in the direction of the shift, i.e.,
no ``downwind'' information is used. For higher order schemes, the
importance of upwinding may be less significant, since the stencils
are large relative to the size of the shift vector. In practise, some
implementations have empirically determined that upwinding by 1 point
at 6th-order is helpful~\cite{Husa:2007rh}. However,
this is not done universally, particularly in conjunction with
8th-order centred differencing~\cite{Lousto:2008dn,
Campanelli:2008nk}.

We have found upwinding to be important in reducing numerical error in
the black hole motion for every order of accuracy we have tried.  The
effect is demonstrated in Fig.~\ref{fig:u8D8-phasing-no-upwinding},
which plots the motion of the black hole punctures for a data set
involving a pair of equal-mass binaries with spins $a_1=-a_2=0.8$
evolved at a relatively low resolution with 8th-order spatial finite
differencing. The results of two evolutions
are plotted, one using fully centred stencils, and the other upwinding
the advection terms with a one-point offset. Whereas the latter
evolution displays the expected inspiral behaviour, at this resolution
the binary evolved with centred advection actually flies apart. The is
purely a result of accumulated numerical error, and at higher resolutions both
tracks can be made to inspiral and merge. Our observation, however, is
that for a given fixed resolution, the one-point offset advection has
a significantly reduced numerical error in the phase as compared to
the fully centred derivatives.

Based on some limited experimentation with larger offsets, we have the
general impression that the one point offset provides the optimal
accuracy for each of the finite difference orders we have tried (4th,
6th, 8th). We do not exclude the possibility that there may be
situations in which the fully centred stencils perform as well as
upwinded advection, however we have not come across a situation where
the latter method performs worse.

As an alternative, we have also tested lower order upwinded
derivatives as a potential scheme which would allow us to maintain a
smaller stencil width. We generally find that the resultant numerical
errors are of the same magnitude or larger than if we had not done the
upwind at all.

We note parenthetically the fact that the off-centering is most
important in the immediate neighbourhood of the black holes, where
the shift has a non-trivial amplitude. It is possible that a scheme
where the stencils are off-centred only on grids where the shift
is larger than some threshold would also be effective, and not
suffer the drawbacks mentioned above over the bulk of the grid. We
have not experimented with such a scheme, however.

%%%%%%%%%%%%%%%%%%%%%%%%%%%%%%%%%%%%%%%%%%%%%%%%%%%%%%%%%%%%%%%%%%%%%%%%%%%%%%
\section{High order finite differencing}
\label{sec:eighth_order}

\begin{figure}
  \begin{center}
    \includegraphics[width=1.\linewidth,clip,trim=15 10 25 0]
      {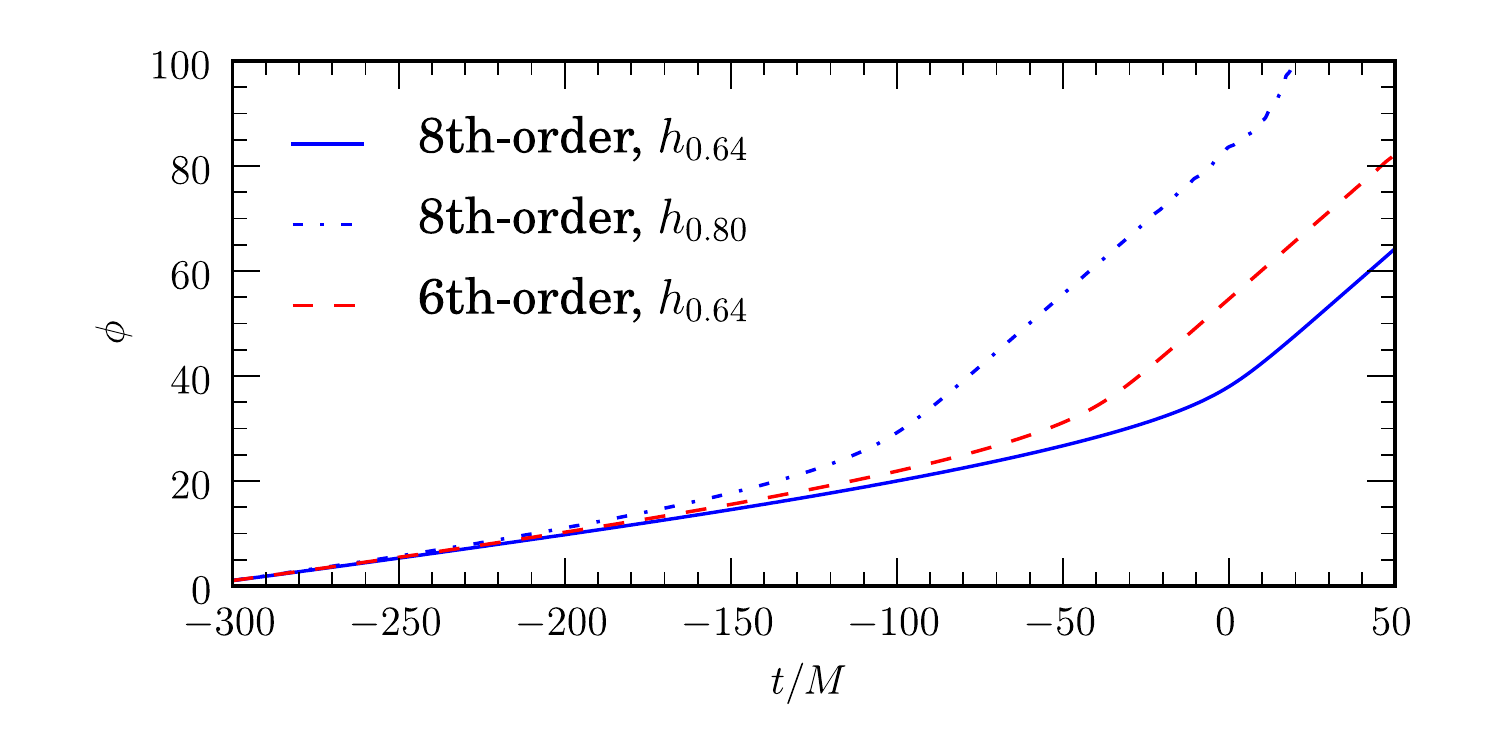}
  \end{center}
  \vspace*{-2\baselineskip}
  \caption{Phase evolution of the $(\ell,m)=(2,2)$ mode $\psi_4$ for
    the aligned-spin model with $a_1=-a_2=0.8$ $h=0.64M$. The
    6th-order case at $h_{0.64}$ has a trajectory between the low
    resolution ($h_{0.80}$) and high resolution ($h_{0.64}$) 8th-order
    evolution.}
  \label{fig:u8D8-phasing-fd6-vs-fd8}
\end{figure}

A recent trend in the implementation of finite difference codes for
relativity has been the push towards higher order spatial derivatives,
It is now common to use 6th or 8th-order stencils. The benefit of
higher order stencils is that the convergence rate can be dramatically
increased, so that a small increase in resolution leads to a large
gain in accuracy.  And while not guaranteed, it is often the
case that for a given fixed resolution, a higher order derivative will
be more accurate, requiring fewer points to accurately represent a
wavelength~\cite{Gustafsson95}.

In moving to high order stencils, there is a trade-off between the
possible accuracy improvements, and the extra computational cost.
High order stencils generally involve two extra floating point
operations per order. Since they require a larger stencil width, they
also incur a cost in communication of larger ghost zones, as well as
requiring wider overlap zones at grid boundaries. In practice, we find
that higher order stencils can also have a more strict Courant limit,
requiring a smaller timestep (and thus more computation to reach a
given physical time). While it is possible to demonstrate a large gain
in accuracy in switching from 2nd to 4th-order operators, there are
diminishing returns in the transition to 6th and higher
order~\cite{Gustafsson95}.

We have experimented with 4th, 6th and 8th-order finite differencing
for the evolution equations. Generally we find that the 8th-order
operators can indeed provide a notable benefit, particularly in the
phase accuracy, at low resolution. In
Fig.~\ref{fig:u8D8-phasing-fd6-vs-fd8}, we plot the phase evolution
for an equal mass model with spins $a_1=-a_2=0.8$. The evolution
covers the last three orbits and ringdown. We find that for this
high-spin case, even over this short duration, a significant dephasing
takes place. Assuming 8th-order convergence, the 6th-order evolution
at the $h_{0.64}$ resolution would be comparable to the 8th-order at
approximately $h_{0.77}$ resolution. We can get some idea of the
relative amount of work required for each calculation by noting there
would be $N=(0.64/0.77)^3$ fewer grid points in the $h_{0.77}$
evolution, but the 8th-order derivatives require $9/7$ times as many
floating point computations for a derivative in one coordinate
direction, and requires a Courant factor which is $0.9$ times that of
the 6th-order run. Taken together, this suggests an 8th-order run at
$h_{0.77}$ would require a factor $0.68$ of the amount of work of the
6th-order case to achieve comparable accuracy. Note that this
computation does not take into account potential additional
communication overhead associated with the wider 8th-order
stencils. But assuming this is not dominant, the conclusion seems to
be that for this level of accuracy, the 6th-order evolution is
somewhat less efficient than the 8th-order version would be.

For a given situation, it may be that these factors change
significantly.  Implementation, and even hardware, details can shift
the balance of costs between various operations. Further, the test
case considered here involves a fairly high spin.  Lower spin models
(such as that considered in the main body of the paper), are accurate
at modest resolutions, and in such cases the 6th-order evolutions may
in fact prove to be relatively more efficient if the accuracy is
already sufficient for a given purpose. On the other hand, if grid
sizes and memory consumption are limiting factors, the 8th-order
operators do give a consistent accuracy benefit for a fixed grid
size. Our expectation, however, is that implementing yet higher order
stencils (for example, 10th-order) may not be justified on the basis
of efficiency.

\begin{figure}
  \begin{center}
    \includegraphics[width=1.\linewidth,clip,trim=25 0 25 0]
      {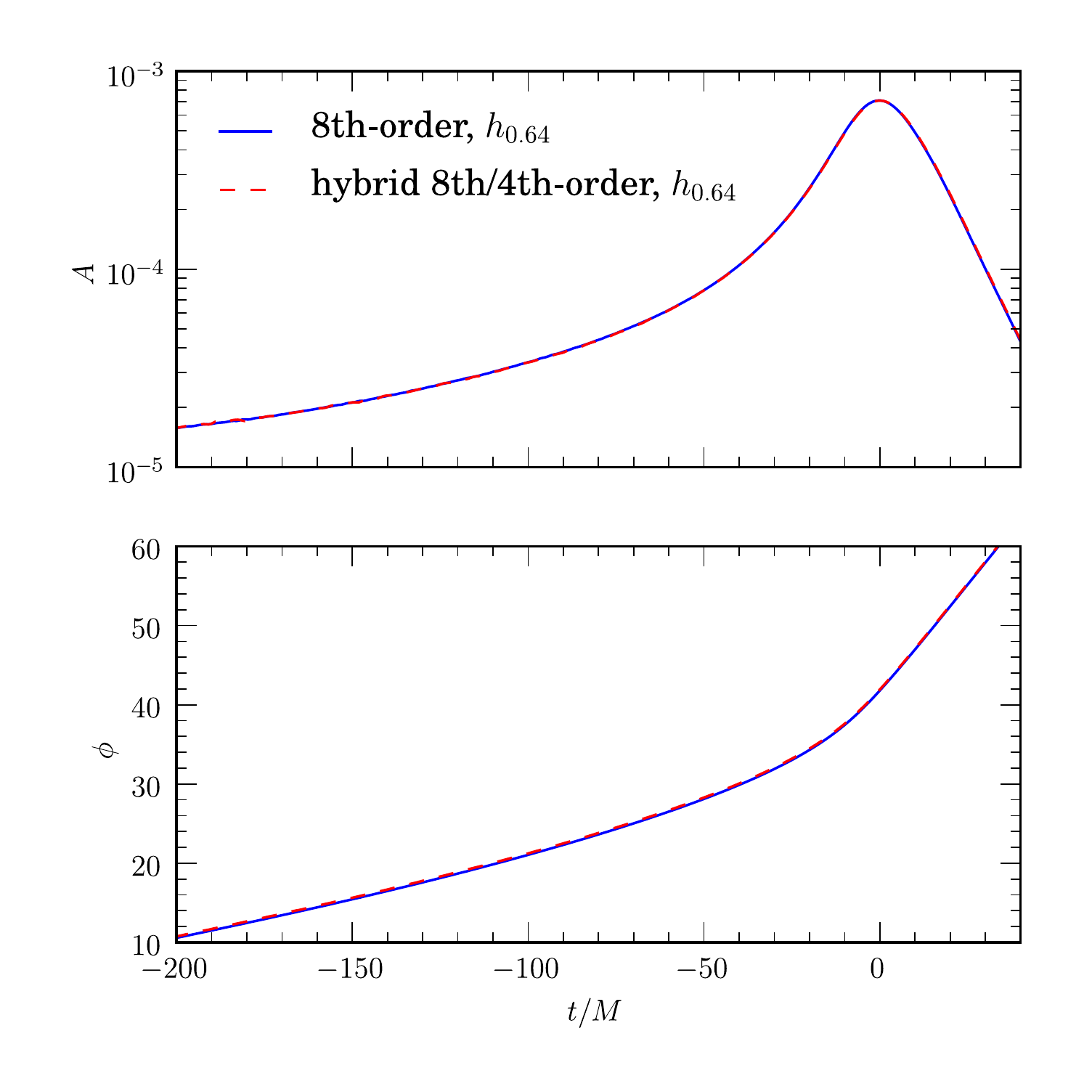}
  \end{center}
  \vspace*{-2\baselineskip}
  \caption{Amplitude and phase evolution of the $(\ell,m)=(2,2)$ mode
    of $\psi_4$ for the equal-mass aligned-spin model, comparing
    8th-order spatial finite differencing with a scheme
    in which 8th-order is used only on the fine meshes surrounding
    the bodies, and 4th-order on the wave-zone grids.}
  \label{fig:u8D8-phasing-hybrid}
\end{figure}

As a final point, we note that the required high-order accuracy
appears to be largely a consequence of the field gradients in the
near-zone, immediately surrounding the black holes. An alternative
scheme, then, could be to apply high-order finite differencing in this
region, while using a lower order (and thus more efficient) scheme in
the wave zone. Results from such a test are displayed in
Fig.~\ref{fig:u8D8-phasing-hybrid}, where we have used 8th-order only on
the finest refinement level, \ie, the mesh surrounding 
the black holes, but 4th-order on all coarser
Cartesian and radial wave-zone grids.  This, in turn, allows for a
slightly less restrictive Courant limit, so that it becomes possible to
run with a slightly larger time-stepping.  The phase evolution of
$\psi_4$ is almost identical to that of the fully 8th-order case, but
the we found that the speed of the run was increased by more than
$25\%$ (similar to that of the fully 6th-order evolution). Further
optimisations, such as decreasing ghost-zone sizes of the 4th-order
grids and consequently the communication overhead, might improve this
further. While the errors and convergence order of this scheme have
not been tested in detail, we suggest it as a potentially quite
effective scheme for the impatient.

%%%%%%%%%%%%%%%%%%%%%%%%%%%%%%%%%%%%%%%%%%%%%%%%%%%%%%%%%%%%%%%%%%%%%%%%%%%%%%
\section{Choice of conformal variable}
\label{sec:conformal_factor}

\begin{figure}
  \begin{center}
    \includegraphics[width=\linewidth,clip,trim=25 0 25 0]
      {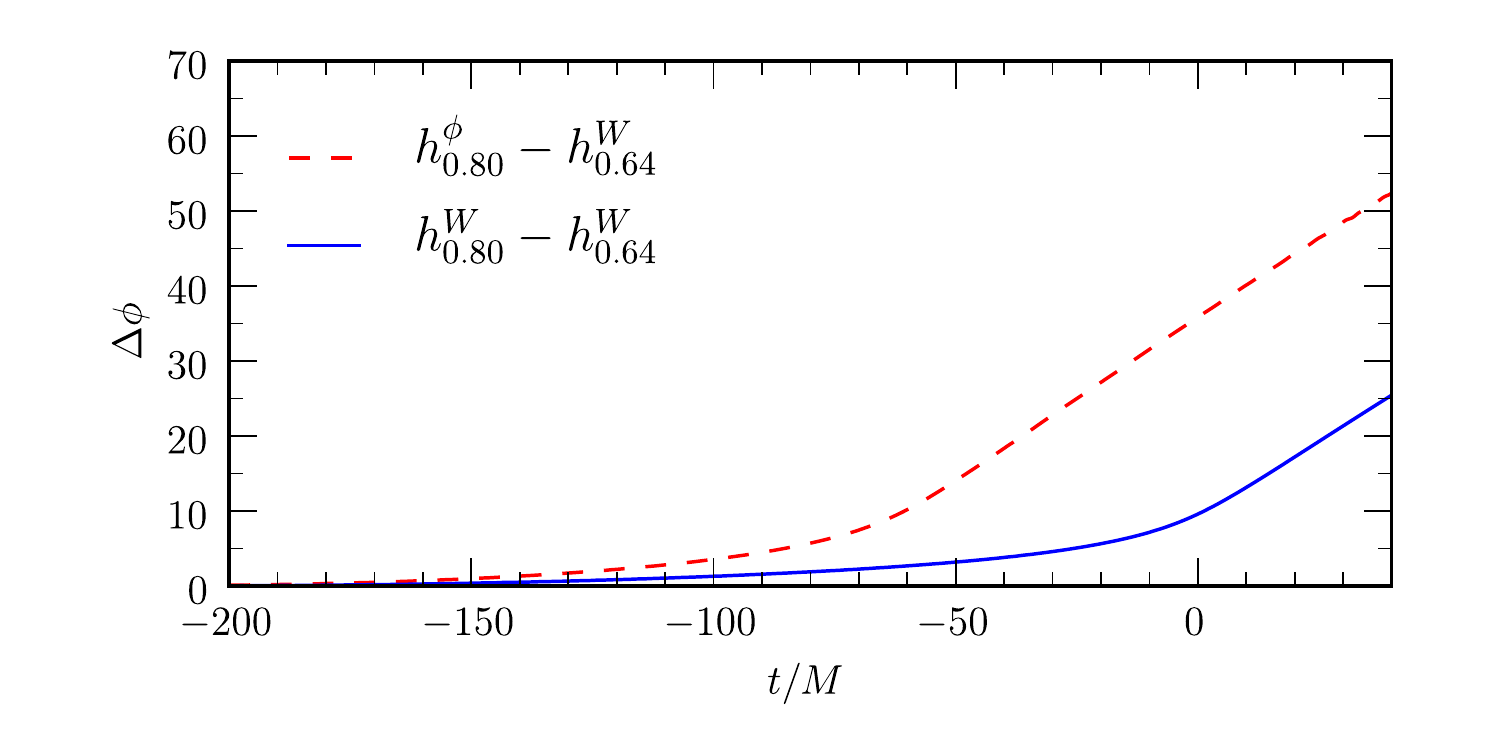}
  \end{center}
  \vspace*{-2\baselineskip}
  \caption{Differences in phase of a spinning configuration with
    resolution $h=0.80M$ and conformal variables $\phi$ and $W$ against a
    simulation with $h=0.64M$ and conformal variable $W$. The dephasing
    is significant as we are on the coarse limit of resolution for
    this particular configuration.}
  \label{fig:u8D8-phasing}
\end{figure}

In Sec.~\ref{sec:evolution}, we have described our implementation
of the BSSNOK evolution system, and note that currently three
variations are in use, based on the use of different variables to
represent the conformal scalar. The original formulation is based on
the use of $\phi := \log \gamma / 12$. An issue with this variable in
the context of puncture evolutions is that it has an $O(\ln r)$
singularity which can lead to large numerical error in finite
differences calculated in the neighbourhood of the puncture.  More
recently, the use of alternative variables $\chi =
\gamma^{-1/3}$~\cite{Campanelli:2005dd} and $W =
\gamma^{-1/6}$~\cite{Marronetti:2007wz} have been proposed as a means
of improving this situation by replacing $\phi$ with variables that
are regular everywhere on the initial data slice.  In terms of the
evolution system outlined in Eqs.~\eqref{eq:bssn}, the $\chi$ and $W$
options correspond to the choices $\kappa=3$ and $\kappa=6$,
respectively.

The influence of this change of variable can be seen in improved phase
accuracy of binary evolutions carried out with either $\chi$ or $W$.
In Fig.~\ref{fig:u8D8-phasing}, we show results from an evolution of
the equal-mass aligned-spin ( $a_1=-a_2=0.8$) test case presented in
the previous appendices, using $\phi$ and $W$ as evolution
variables. Plotted are the phase errors, $\Delta \phi$, between runs
at low resolution, $h_{0.80}$, using both $\phi$ and $W$ with a higher
resolution, $h_{0.64}$, evolution using $W$. The numerical error
associated with the low resolution $\phi$ evolution is significantly
larger than that of the corresponding $W$ evolution.

The reason for this may be related to that of the benefit seen from
upwind advective differences in Appendix~\ref{sec:upwind}. The phase
accuracy of the waveforms is crucially dependent on correctly modelling
the motion of the bodies, and this requires accurate advective
derivatives in the neighbourhood of the punctures. The reduced
numerical error associated with the regular $\chi$ and $W$ variables
is important.

Note that even in the $\phi$ case, numerical error generated at the
puncture seems to be confined to within the horizon. Quantities such
as constraints measured outside the horizon, or the horizon properties
itself, are not significantly affected. However, it seems that a
clear reduction in phase error can be attained through the use of
either the $\chi$ or $W$ variants of BSSNOK, and we have used the
latter for the tests carried out in this paper.

%%%%%%%%%%%%%%%%%%%%%%%%%%%%%%%%%%%%%%%%%%%%%%%%%%%%%%%%%%%%%%%%%%%%%%%%%%%%%%
%%% References
%%%%%%%%%%%%%%%%%%%%%%%%%%%%%%%%%%%%%%%%%%%%%%%%%%%%%%%%%%%%%%%%%%%%%%%%%%%%%%
\bibliographystyle{apsrev-nourl}
\bibliography{aeireferences}

\end{document}